\documentclass[apjl]{emulateapj}
\usepackage{epsfig}
\usepackage{verbatim}
\usepackage{natbib}
\usepackage{apjfonts}

\def\lae{\mathrel{<\kern-1.0em\lower0.9ex\hbox{$\sim$}}}
\def\gae{\mathrel{>\kern-1.0em\lower0.9ex\hbox{$\sim$}}}
\def\deg{^{\circ}}

\begin{document}

\shortauthors{TREMBLAY ET AL.}
\shorttitle{STAR FORMATION IN 3C~236}

\title{Episodic Star Formation Coupled to Reignition of Radio Activity in 3C~236}

\slugcomment{Accepted for Publication in ApJ}

\author{Grant R.~Tremblay}
\author{Christopher P.~O'Dea}
\author{Stefi A.~Baum}
\affil{Rochester Institute
of Technology, One Lomb Memorial Drive, Rochester, NY 14623, USA; grant@astro.rit.edu}
\author{Anton M.~Koekemoer}
\author{William B.~Sparks}
\affil{Space  Telescope Science  Institute, 3700  San Martin
Drive, Baltimore, MD 21218, USA}
\and
\author{Ger de Bruyn}
\author{Arno P.~Schoenmakers}
\affil{Stichting Astronomisch Onderzoek in Nederland, 
P.O.~Box 2, 7990 AA Dwingeloo, The Netherlands}

\begin{abstract}
We present  {\it Hubble Space Telescope} ACS  and STIS FUV/NUV/optical
imaging of the radio galaxy 3C~236, whose relic $\sim 4$ Mpc radio jet
lobes and inner 2 kpc CSS radio source are evidence of multiple epochs
of AGN  activity. Consistent with  previous results, our  data confirm
the presence of four bright knots  of FUV emission in an arc along the
edge of the inner circumnuclear  dust disk in the galaxy's nucleus, as
well as FUV emission cospatial  with the nucleus itself.  We interpret
these to  be sites  of recent or  ongoing star formation.   We present
photometry of  these knots,  as well as  an estimate for  the internal
extinction in the source  using line ratios from archival ground-based
spectroscopy.   We estimate  the ages  of the  knots by  comparing our
extinction-corrected  photometry  with  stellar  population  synthesis
models. We  find the four  knots cospatial with  the dusty disk  to be
young, of order $\sim10^7$ yr old. The FUV emission in the nucleus, to
which  we do not  expect scattered  light from  the AGN  to contribute
significantly, is likely due to an episode of star formation triggered
$\sim10^9$ yr  ago.  We argue that  the young $\sim10^7$  yr old knots
stem from  an episode of star  formation that was  roughly coeval with
the event resulting in reignition  of radio activity, creating the CSS
source.  The $\sim10^9$ yr old  stars in the nucleus may be associated
with the  previous epoch  of radio activity  that generated the  4 Mpc
relic source, before being cut off by exhaustion or interruption.  The
ages of  the knots,  considered in the  context of both  the disturbed
morphology of the nuclear dust and the double-double morphology of the
``old''  and   ``young''  radio  sources,  present   evidence  for  an
AGN/starburst  connection that  is  possibly episodic  in nature.   We
suggest that  the AGN fuel supply  was interrupted for  $\sim 10^7$ yr
due to a  minor merger event and has now  been restored. The resultant
non-steady flow of gas in the  disk is likely responsible for both the
new  episode of infall-induced  star formation  and also  the multiple
epochs of radio activity.
\end{abstract}


\section{Introduction}

\begin{figure*}
\plottwo{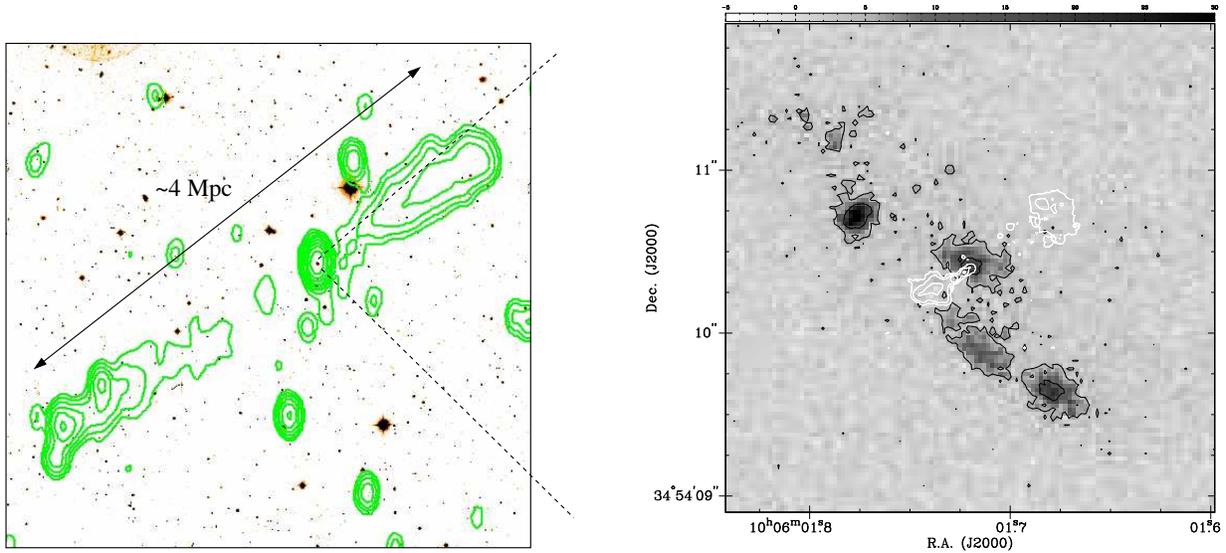}{csssource.ps}
\caption{The  two radio sources  associated with  3C~236. (a)  326 MHz
  WSRT  radio  contours  (in  green,  from  \citealt{mack97})  of  the
  ``relic'' radio  emission associated with 3C~236,  overlayed on SDSS
  imaging of the same region of  sky.  The deprojected size of the jet
  is $\sim 4.5$ Mpc, making it  the largest known radio galaxy and one
  of the  largest objects in the  universe.  (b) Global  VLBI 1.66 GHz
  radio contours  of the  central 2 kpc  CSS (``young'')  radio source
  from \citet{schilizzi01}, overlayed on {\it HST}/STIS NUV imaging of
  the star forming knots described by \citet{odea01}.  The jet axes of
  the Mpc- and kpc- scale radio sources are aligned on nearly the same
  position angle.  }
\label{fig:doubledouble}
\end{figure*}

Galaxies  occupy  a heavily  bimodal  distribution in  color-magnitude
space, wherein young,  predominantly disk-dominated galaxies reside in
a  `blue  cloud'  and  evolve  onto  a  characteristically  quiescent,
bulge-dominated `red  sequence' (e.g., \citealt{bell04,faber07}).  The
underdensity  of  galaxies  in  the `green  valley'  separating  these
populations  implies   that  cloud-to-sequence  evolution   is  swift,
requiring  a cessation  of star  formation  more rapid  than would  be
expected  in  passively  evolving systems  (e.g.,  \citealt{cowie96}).
Quasar-  and   radio-mode  feedback  models  have   been  proposed  as
mechanisms by which star formation may be truncated by the heating and
expulsion of gas \citep{silk98,hopkins05,croton06,schawinski06}, as it
is now  known that  quasar activity was  two orders of  magnitude more
common  at  redshifts  $z\sim2$   than  at  the  present  time  (e.g.,
\citealt{schmidt91}).   This, considered in  the context  of declining
star  formation   rates  in   massive  galaxies  at   $z\sim2$  (e.g.,
\citealt{perezgonzalez07}),  along with  the  emerging consensus  that
most  populations of galaxies  harbor quiescent  black holes  at their
centers (hereafter BHs, e.g,  \citealt{kormendy95}), has given rise to
questions of whether all bright galaxies go through one or more active
phases  (e.g., \citealt{haehnelt93,cavaliere89}).   In  this scenario,
the  quenching of  star formation  via feedback  from  active galactic
nuclei (AGN)  may be one of  the primary drivers  of cosmic downsizing
(e.g., \citealt{cowie96,scannapieco05}, and references therein).

The relationship between the AGN duty cycle and the regulation of host
galaxy stellar evolution  is far more complicated, however,  as it can
play competing  roles at successive stages of  galactic evolution. AGN
activity has been associated not only with quenching star formation on
large scales,  but also triggering  it via ISM cloud  compression from
the propagating relativistic jets associated with radio galaxies (e.g,
the             so-called             ``alignment            effect'',
\citealt{rees89,baum89a,mccarthy93,best00,  privon08}).   Moreover, it
is  natural  to  expect  a  correlated (but  not  necessarily  causal)
relationship  between  AGN activity  and  star  formation.  The  tight
relationship between BH mass and host galaxy bulge velocity dispersion
\citep{magorrian98,ferrarese00,gebhardt00} implies  that the growth of
the   BH   and  the   galaxy   bulge   are   tightly  coupled   (e.g.,
\citealt{kauffmann00,ciotti01}).   It   is  therefore  expected  that,
throughout the  process of  hierarchical galaxy formation,  gas infall
due to major mergers or  tidal stripping from a gas-rich companion can
fuel not  only AGN,  but also  the growth of  the host  galaxy stellar
component via  infall-induced starbursts (e.g., \citealt{dimatteo05}).
A significant fraction ($\sim 30\%$) of nearby powerful radio galaxies
exhibit  evidence  of  infall-induced  starbursts  near  their  nuclei
\citep{smith89,allen02,baldi08,tremblay09},    suggesting   that   the
phenomenon is both common and  comperable to the lifetime of the radio
source    ($\sim10^7-10^8$   yr,   e.g.,    \citealt{parma99}).    The
AGN/starburst connection  is therefore likely real  and fundamental to
galaxy  evolution itself, and  its characterization  has been  a major
pursuit      of       the      past      two       decades      (e.g.,
\citealt{barnes91,silk98,fabian99,dimatteo05,hopkins05,springel05,silverman08,quillen08a}).

A key  discriminant in understanding  the nature and evolution  of the
AGN/starburst  connection may be  found in  some radio  galaxies whose
morphology  is clear  evidence for  multiple epochs  of  AGN activity.
Several   such  examples   have   been  observed   (e.g.,  3C~219   --
\citealt{bridle86, clarke92}, 0108+388  -- \citealt{baum90}), and have
come  to constitute a  new class  of ``double-double''  radio sources,
representing  $\sim 5-10$\%  of predominantly  large ($>1$  Mpc) radio
galaxies       (e.g.,      \citealt{schoenmakers00a,schoenmakers00b}).
Double-doubles  are   characterized  by  outer   (`older')  and  inner
(`younger') radio sources propagating outwards amidst the relic of the
previous epoch  of activity.   This apparently repetitive  activity is
thought  to  be a  consequence  of the  AGN  fuel  supply having  been
interrupted,  whether by  exhaustion, smothering,  or  disturbance, at
some  time in the  past \citep{baum90}.   This scenario  is consistent
with     models     of     radio     galaxy     propagation     (e.g.,
\citealt{kaiser00,brocksopp07}).   The  relative  ages  of  the  radio
sources (and therefore the timescale over which the engine was cut off
and re-ignited) can be estimated  using size estimates from radio maps
coupled with a dynamical model  for the jets and radio spectral energy
distributions      (SEDs)     of      the      radiating     electrons
\citep{schoenmakers00a,schoenmakers00b,odea01}.

\subsection{An important test case: 3C~236}

The  nearby  ($z=0.1005$)  double-double  radio galaxy  3C~236  is  an
important test case in studies of the AGN/starburst connection, and is
the basis of  both this paper and a  previous study by \citet{odea01}.
3C~236 is a powerful  double-double with a relic edge-brightened FR~II
\citep{fanaroff74}  radio  source   whose  deprojected  linear  extent
exceeds 4 Mpc,  making it the second largest  known radio galaxy (only
J1420-0545  is larger,  \citealt{machalski07}),  and even  one of  the
largest objects in the  universe \citep{schilizzi01}.  Its inner young
Compact  Steep  Spectrum  (CSS)   source,  whose  apparent  origin  is
cospatial  with  the  nucleus,  is   only  2  kpc  in  extent  and  is
morphologically reminiscent of a young classical double.  Anecdotally,
3C~236 was  initially classified  as a pure  CSS source before  it was
associated years later with  the massive relic FR~II source (R.~Laing,
private communication).  The jet propagation axes of both the Mpc- and
kpc- scale sources  are aligned to within $\sim10\deg$  of one another
(as projected on the  sky).  See Fig.~\ref{fig:doubledouble} for radio
contour overlays  of both sources, using 326  MHz Westerbork Synthesis
Radio  Telescope  (WSRT)  and  1.66  GHz  Global  Very  Long  Baseline
Interferometry  (VLBI) Network radio  mapping from  \citet{mack97} and
\citet{schilizzi01} for the relic and CSS sources, respectively.

\begin{figure}
\plotone{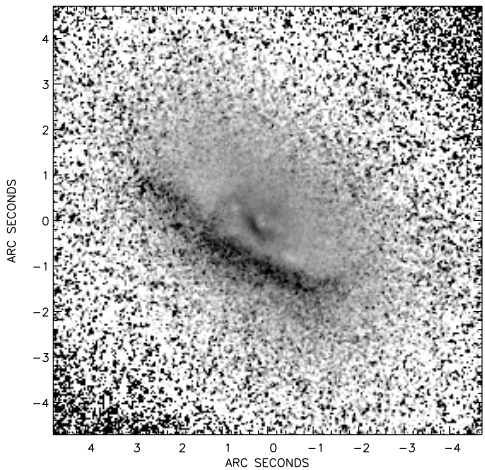}
\caption{$1.6$  $\mu$m /  0.7 $\mu$m  colormap of  the outer  lane and
  inner dusty disk in the nucleus of 3C~236, made via division of {\it
    HST}/NICMOS   and    WFPC2   data   from    \citet{martel99}   and
  \citet{madrid06}, respectively.  This absorption map  was originally
  presented in \citet{tremblay07}. }
\label{fig:colormap}
\end{figure}

In addition  to its  rare radio morphology,  3C~236 is also  unique in
that its  nuclear dust  complex is made  up of an  inner circumnuclear
dusty  disk that is  somewhat misaligned  with an  apparently separate
outer dust lane  \citep{martel99,dekoff00,tremblay07}.  The total dust
mass in the complex is estimated to be $\sim 10^7$ M$_\odot$, based on
{\it  Hubble Space  Telescope} ({\it  HST}) absorption  maps  and IRAS
luminosities \citep{dekoff00}.   In Fig.~\ref{fig:colormap} we present
a  $1.6$ $\mu$m  /  0.7 $\mu$m  absorption  map of  the dust  complex,
originally presented  in \citet{tremblay07}  and made via  division of
{\it   HST}/NICMOS   and   WFPC2   data  from   \citet{martel99}   and
\citet{madrid06}, respectively.


\begin{deluxetable*}{ccccccccc}
\tablecaption{Summary of Observations of 3C~236}
  \tablewidth{0pc}
  \tablehead{
    \colhead{Observatory} &
    \colhead{Instrument} &
    \colhead{Aperture} &
    \colhead{Filter/Config.} &
    \colhead{Waveband/Type} &
    \colhead{Exp. Time [Orbits]} &
    \colhead{Reference}  &
    \colhead{Obs. Date} &
    \colhead{Comment} \\
    \colhead{(1)} & \colhead{(2)} & \colhead{(3)} &
    \colhead{(4)} & \colhead{(5)} & \colhead{(6)} &
    \colhead{(7)} & \colhead{(8)} & \colhead{(9)} }
  \startdata
\cutinhead{{\sc New Observations}}
{\it HST}  & ACS & HRC   &  F330W  & $U$-band Imaging&  2516s [1] &  {\it HST} 9897 & 21 Oct 2003 & SF Knots \\
{\it HST}  & ACS  & HRC  &  F555W & $V$-band Imaging &  2612s [1] & {\it HST} 9897   & 22 Oct 2003 & Dust Lanes   \\
{\it HST}  & ACS  & SBC  &  F140LP &   FUV Imaging & 6900s [3]  & {\it HST} 9897 & 21 Oct 2003 & SF Knots \\
{\it HST}  & STIS & NUV-MAMA    &  F25SRF2  & NUV Imaging & 2520s[1]  &  {\it HST} 9897 & 19 Oct 2003 & SF Knots \\
\cutinhead{{\sc Archival Observations}}
{\it HST}  & WFPC2 & PC1    &  F702W  & $R$-band Imaging&  4$\times$140s &  {\it HST} 5476  & 7 May 1995 & Galaxy \\
{\it HST}  & WFPC2 & PC1    &  F555W  & $V$-band Imaging&  2$\times$300s &  {\it HST} 6384  & 12 Jun 1996 & Galaxy \& Dust \\
{\it HST}  & STIS &  NUV-MAMA   &  F25SRF2  & NUV Imaging &  1440s &  {\it HST} 8275  & 03 Jan 1999 & SF Knots \\
{\it HST}  & NICMOS &  NIC2-FIX   &  F160W  & NIR Imaging &  1152s &  {\it HST} 10173  & 02 Nov 2004  & Host Isophotes\cr
%
  \enddata
  \tablecomments{
A summary of the new and archival observations used in our analysis. 
    (1) Facility name;
    (2) instrument used for observation;
    (3) configuration of instrument used;
    (4) filter used;
    (5) corresponding waveband and specification of whether the observation was imaging or spectroscopy;
    (6) exposure time (if the observatory is {\it HST}, the corresponding number of orbits also appears in brackets);
    (7) corresponding reference for observation. If the observatory is {\it HST}, the STScI-assigned program number is listed;
    (8) date of observation;    
    (9) comment specific to observation. 
}
\label{tab:tab1}
\end{deluxetable*}

\begin{figure*}
\plottwo{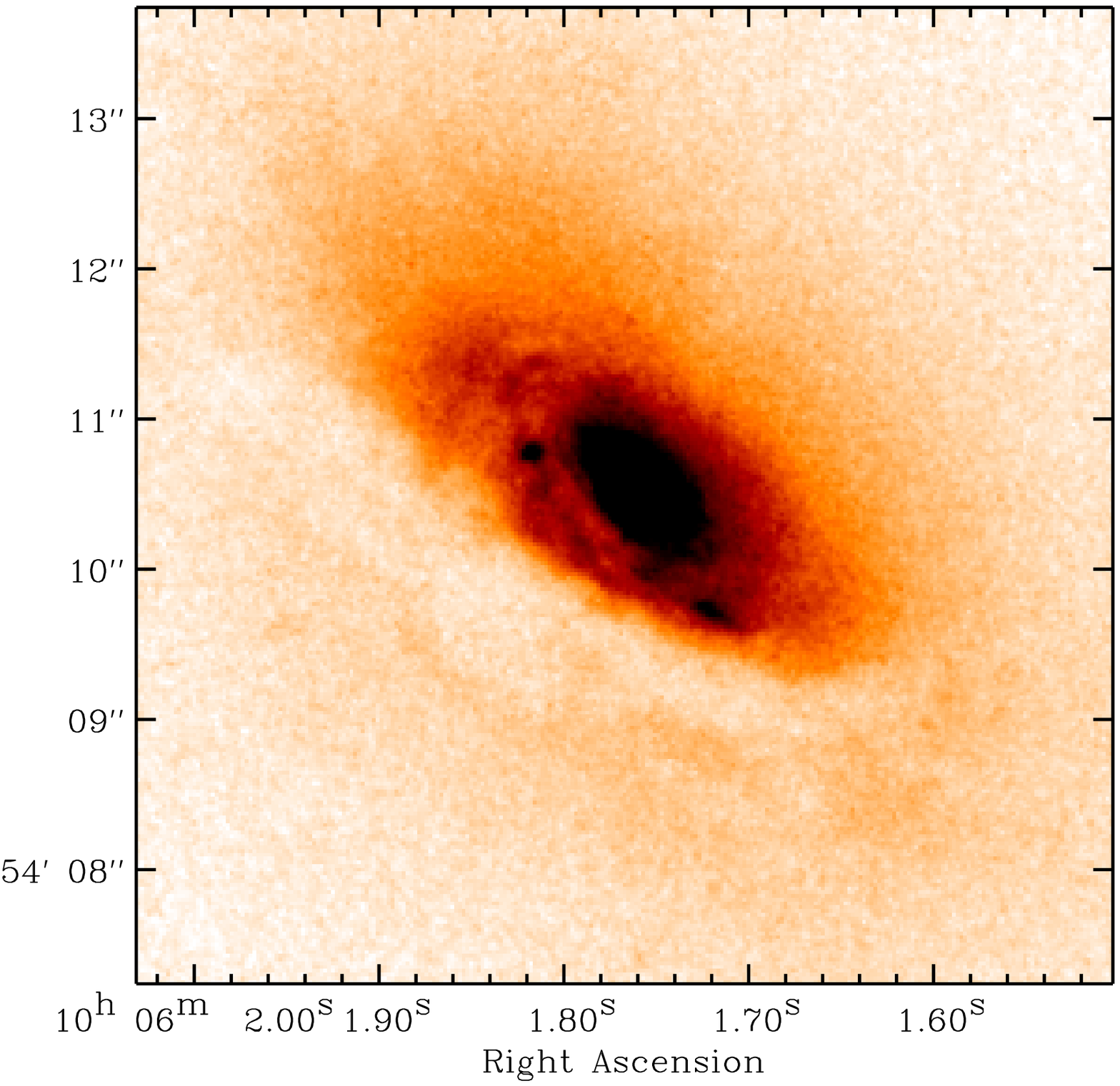}{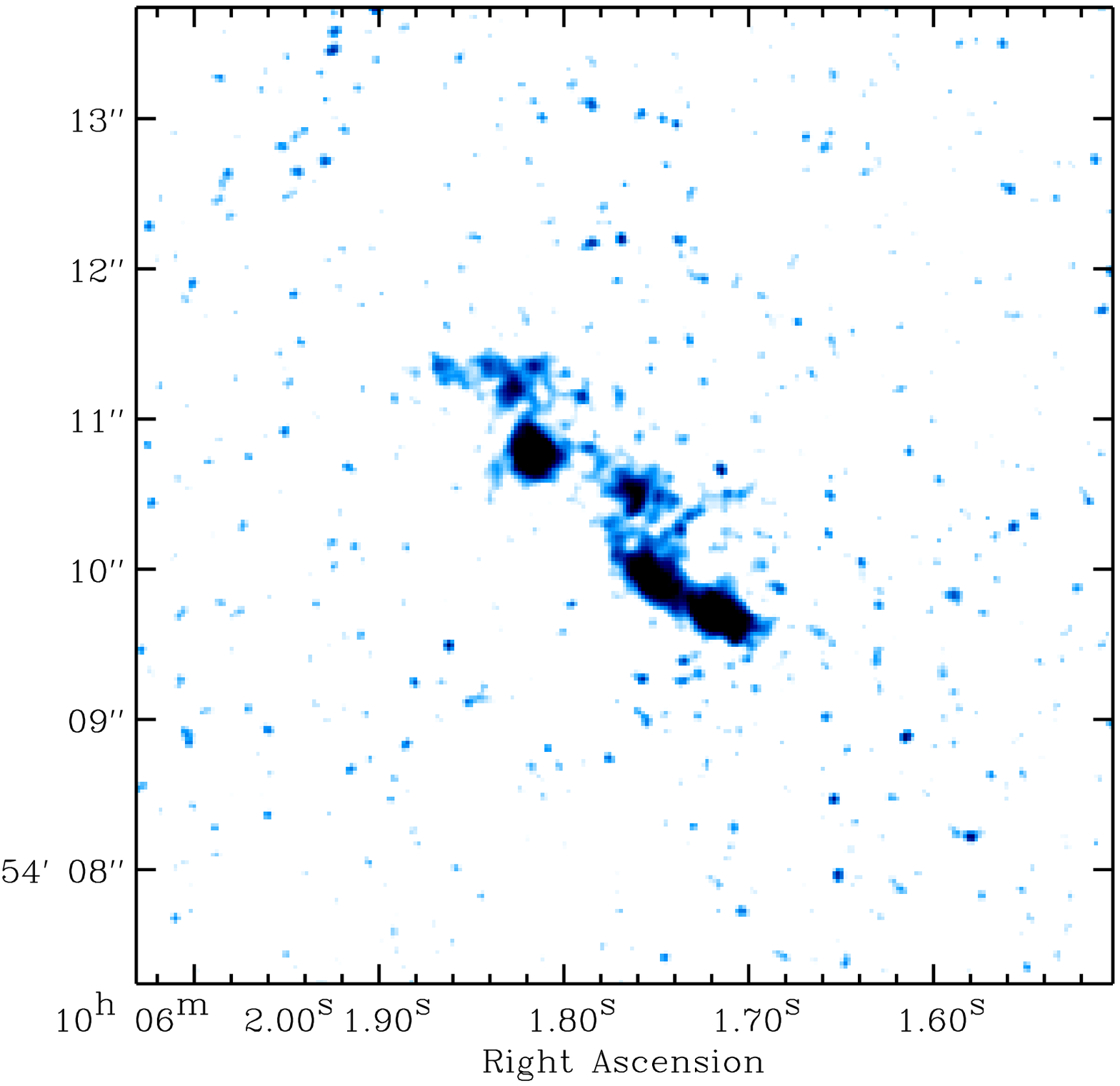}
\caption{({\it  left}) 2500  s  $V$-band exposure  of  the nucleus  of
  3C~236, using {\it HST}/ACS HRC with the broadband F555W filter. The
  outer dust lane  is seen in white, while three of  the four knots of
  star formation are  seen to the south and west  of the nucleus along
  the inner dusty  disk, whose position angle is  slightly offset from
  that of  the outer.({\it right})  combined 6900 s {\it  HST}/ACS SBC
  FUV  (F140LP) image  of the  star forming  knots observed  along the
  inner dust structure of 3C~236.   The image has been smoothed with a
  two  pixel Gaussian  kernel. North  is up,  east is  left.   The two
  images  are  on the  same  scale.  At  a  redshift  of $z\sim  0.1$,
  1\arcsec corresponds to $\sim 1.8$ kpc.}
\label{fig:acs}
\end{figure*}

The work by  \citet{odea01} studied {\it HST} NUV  and optical imaging
of  the central  few  arcsec of  3C~236,  finding four  knots of  blue
emission  arranged in  an  arc along  the  dust lane  in the  galaxy's
nucleus.  Their original NUV data is presented in greyscale with black
contours  in  Fig.~\ref{fig:doubledouble}({\it  b}).  The lack  of  an
obvious  spatial relationship  between the  knots and  the  CSS source
suggests   that  the   starbursts  are   infall-induced   rather  than
jet-induced. \citet{martel99} had also  detected the knots of emission
in their {\it HST} $R$-band imaging,  albeit to a lesser degree as the
knots  are  very  blue.    \citet{odea01}  used  their  photometry  in
comparison with stellar population  synthesis models to estimate upper
limits to  the ages  of the individually  resolved star  forming knots
(seen  in Fig.~\ref{fig:doubledouble}{\it  b}).  They  found disparate
ages  between the  clumps of  emission, finding  two to  be relatively
young with  ages of  order $\sim  10^7$ yr, while  the other  two were
estimated at  $\sim 10^8 - 10^9$  yr old, comparable  to the estimated
age  of the  giant  relic radio  source.   \citet{odea01} argued  that
3C~236  is an  ``interrupted''  radio galaxy,  and  has undergone  two
starburst episodes  approximately coeval with the two  epochs of radio
activity  observed  on Mpc-  and  kpc-  scales.   That work  motivated
follow-up  observations  with  {\it  HST} at  higher  sensitivity  and
spatial resolution, the results of which we present in this paper.

We organize  this work as follows.   In section 2 we  describe the new
and archival data  presented in this paper, as  well as the associated
data  reduction. In  section 3  we  present our  results, including  a
comparison of our photometry with stellar population synthesis models,
following (in the interests of consistency) the analysis strategy used
in \citet{odea01}.  We  discuss our results in section  4, focusing on
the  role  played  by  the  AGN/starburst connection  in  the  special
test-case environment  of 3C~236.  We summarize this  work and provide
some concluding  remarks in section  5.  Throughout this paper  we use
$H_0   =  71$  km   s$^{-1}$  Mpc$^{-1}$,   $\Omega_M  =   0.27$,  and
$\Omega_{\Lambda} = 0.73$.

\section{Observations \& Data Reduction}

We present {\it Hubble Space Telescope} observations of the nucleus of
3C~236 obtained as  part of the Cycle 12 GO program  9897 by O'Dea and
collaborators.  These  consist of $U$-  and $V$-band imaging  with the
Advanced Camera  for Surveys (ACS) High Resolution  Channel (HRC), FUV
imaging with the  ACS Solar Blind Channel (SBC),  and NUV imaging with
the Space  Telescope Imaging  Spectrograph (STIS).  Below  we describe
the specifics  of each  new observation presented  in this  paper, and
provide a  brief description of the archival  imaging and spectroscopy
that we also include in our analysis. We also describe the steps taken
to  reduce the  data.  A  summary of  the new  and archival  {\it HST}
observations  utilized  in  this  work   can  be  found  in  in  Table
\ref{tab:tab1}.

\subsection{Cycle 12 ACS and STIS imaging}

Our  {\it   HST}/ACS  observations   were  designed  to   enable  high
sensitivity multicolor photometry allowing  for construction of an SED
of the  blue knots previously observed in  3C~236 \citep{odea01}.  The
FUV, NUV,  $U$-, and $V$-bands were  chosen so as  to provide multiple
constraints on  a young blue  stellar population, while  also enabling
consistency checks and estimates on the amount of intrinsic reddening.
At the redshift of 3C~236 ($z  = 0.1005$), the $U$- and $V$- bands are
located  just   blue-ward  and  red-ward  of  the   4000  \AA\  break,
respectively.  Exposure  times were chosen to permit  detection of the
knots at adequate signal-to-noise (S/N)  over a range of possible ages
and intrinsic properties.  Three orbits (6900 s) were obtained for the
SBC  FUV image using  the F140LP  long pass  filter.  We  obtained one
orbit each ($\sim 2500$ s) for  the HRC $U$- and $V$-band images using
the F330W and F555W filters, respectively. The F140LP long pass filter
on  ACS  SBC ranges  from  $\sim1350$ \AA\  (with  a  hard cutoff)  to
$\sim2000$  \AA\  with  a  pivot  wavelength of  1527  \AA.   The  SBC
Multianode  Microchannel  Array (MAMA)  has  a  spatial resolution  of
$\sim$0\farcs034 $\times$ 0\farcs030 per  pixel and a nominal field of
view   of  34\farcs6$\times$30\farcs1.   The   SBC  acheives   a  peak
efficiency of 7.5$\%$ at 1250 \AA. The F330W and F555W filters on ACS
HRC have  central wavelengths (filter  widths) of 3354 (588)  \AA\ and
6318  (1442)  \AA\,  respectively.   The  HRC has  a  pixel  scale  of
0\farcs028  $\times$0\farcs025 per  pixel  and its  field  of view  is
29\arcsec$\times$26\arcsec. It reaches a  peak efficiency of 29$\%$ at
$\sim 6500$ \AA.

We  have also  obtained one  orbit  (2520 s)  of NUV  imaging with  the
Cs$_2$Te  MAMA detector  on STIS.   The F25SRF2  filter has  a central
wavelength  of  2320  \AA\ and  a  FWHM  of  1010 \AA,  which  permits
geocoronal  [{\sc o~i}]$\lambda 1302+1306$  \AA\ contamination  in its
bandpass,  though its  contribution  is far  lower  than the  detector
background and  is not  expected to affect  our results.   The F25SRF2
cutoff does  not permit geocoronal Ly$\alpha$  emission.  The NUV-MAMA
has   a  pixel  scale   of  0\farcs024   and  a   field  of   view  of
25\arcsec$\times$25\arcsec.

\subsection{Archival data}

In this paper  we make use of archival  {\it HST} Wide-Field Planetary
Camera  2  (WFPC2) imaging  in  $V$-  and  $R$-band (F555W  and  F702W
filters, respectively)  obtained as part of the  3CR snapshot programs
by Sparks  and collaborators \citep{dekoff96,mccarthy97,martel99}.  We
also  use the  1440 s  STIS NUV-MAMA  F25SRF2 image  of  3C~236, which
formed  the basis  of the  study by  \citet{odea01}, and  was formally
presented as  part of a  data paper by \citet{allen02}.   The $H$-band
image obtained with the {\it HST} Near-Infrared Camera and Multiobject
Spectrograph   (NICMOS2)   in   SNAP   program  10173   (PI:   Sparks,
\citealt{madrid06,tremblay07,floyd08})  was  used for  fits  to  the host  galaxy
isophotes. Imaging and spectroscopy  from the Sloan Digital Sky Survey
(SDSS) is also used \citep{york00,adelman-mccarthy08}.

\begin{figure*}
\plottwo{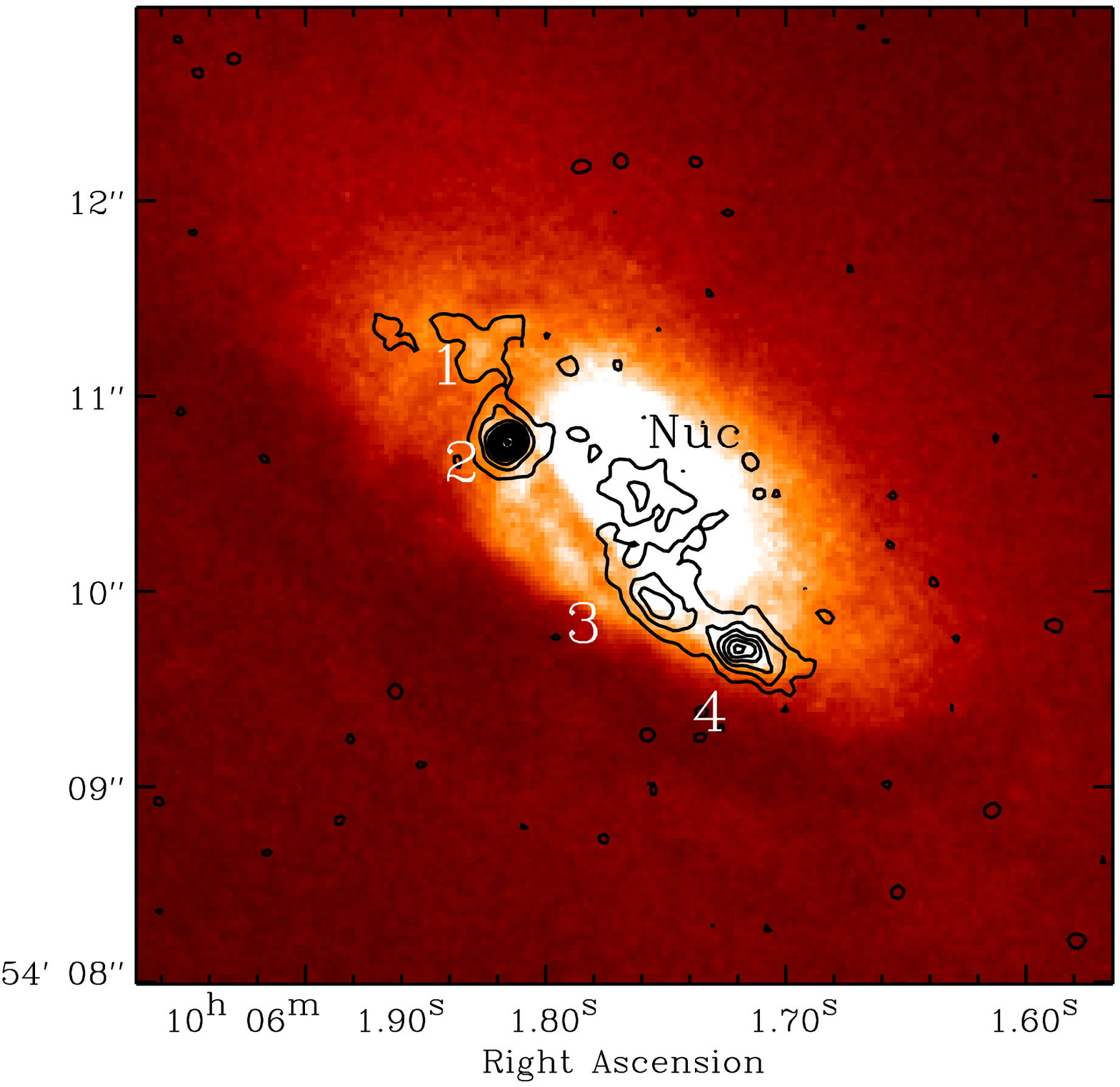}{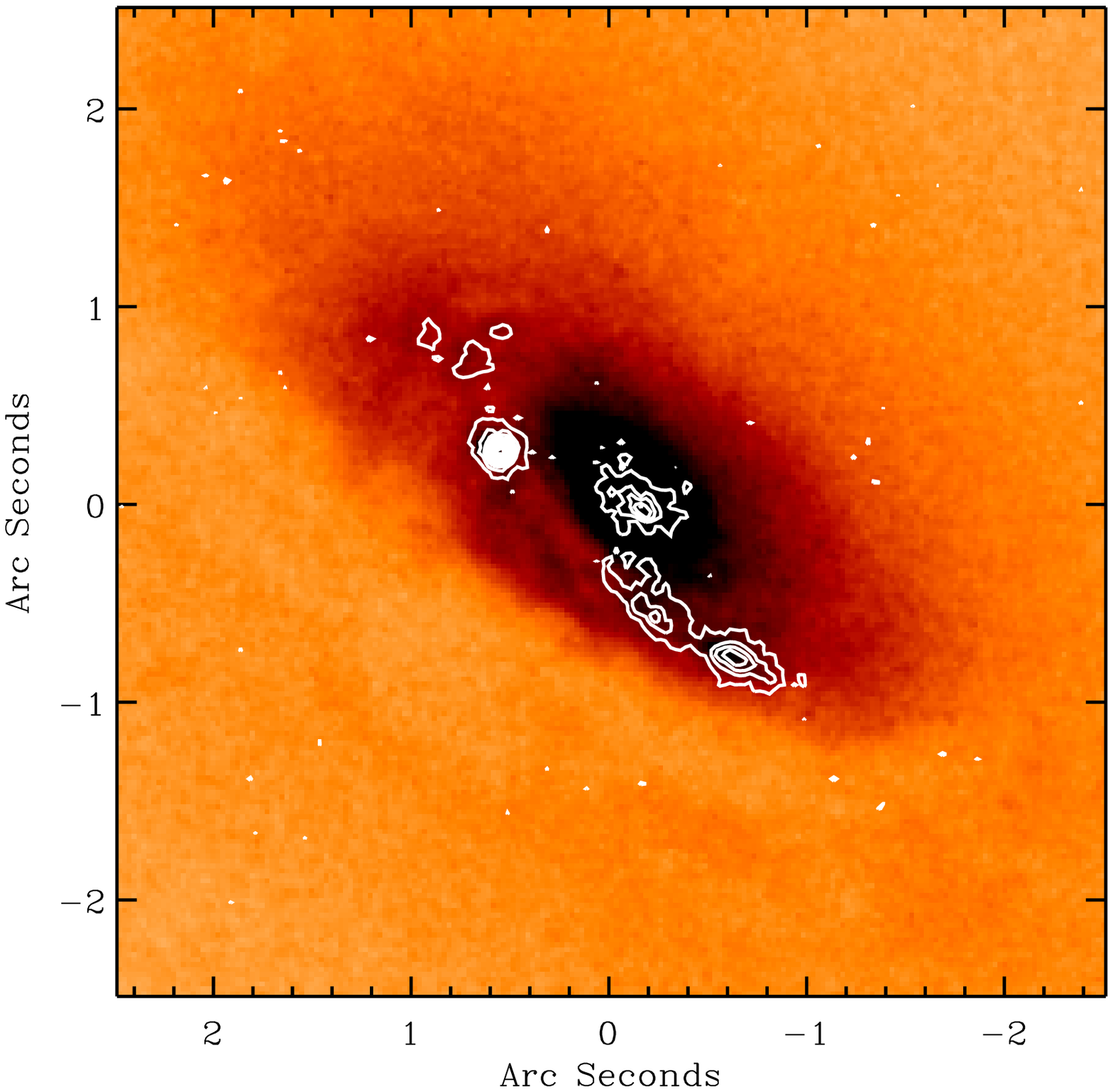}
\caption{({\it a}) ACS HRC $V$-band image of the nucleus of 3C~236 with
  ACS  SBC FUV  contours (smoothed  with a  two pixel  Gaussian).  The
  knots of  star formation  along the inner  dust structure  have been
  labeled  so as  to  be consistent  with  \citet{odea01}.  That  work
  detected knots  ``3'' and ``4''  as two distinct compact  regions of
  emission  in  their STIS  MAMA  NUV  imaging,  while we  detect  one
  continuous  filament of emission  in our  more sensitive  and higher
  spatial resolution FUV imaging.   So as to maintain consistency with
  \citet{odea01} while  making this  clear we refer  to this  patch as
  ``knot  $3+4$''  throughout this  paper.   ({\it  b})  The same  HRC
  $V$-band image  (with colors inverted to better  emphasize the outer
  dust  lane) with  STIS  S25SRF2  NUV contours.   The  knots of  star
  formation are  nearly morphologically identical in both  the FUV and
  NUV,  though the higher  sensitivity of  ACS is  evident in  the FUV
  contours at left, which  highlight lower surface brightness features
  in the knots.  Both panels are aligned and on  the same scale.  East
  is  left,  north  is  up.   At  the  distance  of  3C~236,  1\arcsec
  corresponds to 1.8 kpc.  }
\label{fig:hrccontour}
\end{figure*}

\subsection{Data reduction}

The previously unpublished {\it HST} data presented in this paper were
reduced  using the standard  On-the-Fly Recalibration  (OTFR) pipeline
provided  as part  of  the Multi-mission  Archive  at Space  Telescope
(MAST).   For   ACS,  the   OTFR  pipeline  combines   calibrated  and
flatfielded dithered exposures using the \texttt{multidrizzle} routine
with  the default  parameters.  The  task calculates  and  subtracts a
background sky  value for each  exposure, searches for  additional bad
pixels not already flagged in the data quality array, and drizzles the
input  exposures into  outputs that  are shifted  and  registered with
respect to one another.  From  these drizzled exposures a median image
is created, which is then compared with original input images so as to
reject cosmic  rays on the  drizzled median.  More information  on the
specifics    of     \texttt{multidrizzle}    can    be     found    in
\citet{koekemoer02}.

We have  not performed post-pipeline  processing on the  two one-orbit
ACS HRC  drizzled images ($U$- and  $V$-band), as for  the purposes of
this  work we  are concerned  primarily with  high  surface brightness
emission near the center of the  galaxy, and the results from the OTFR
pipeline were  sufficiently free of cosmic  rays and hot  pixels to be
deemed  ``science  ready''.  For  the  three-orbit  ACS  SBC data,  we
combined  the  three   individual  calibrated  and  flatfielded  files
manually using \texttt{multidrizzle} with the default parameters.  The
single 6900s output image was left un-rotated with respect to North so
as to  avoid associated pixel interpolation errors  in our photometry.
Information regarding  the reduction of the archival  data utilized in
this paper can be found in the appropriate references cited in section
2.2.

\section{Results}

In Fig.~\ref{fig:acs} we present the ACS/HRC $V$-band (at left in red)
and  ACS/SBC FUV  (at right  in blue)  images in  panels (a)  and (b),
respectively. The two  images are aligned and on  the same scale, with
east  left  and  north  up.   At  the  redshift  of  3C~236,  1\arcsec
corresponds to $\sim  1.8$ kpc.  In Fig.~\ref{fig:hrccontour}({\it a})
we present the same data in Fig.~\ref{fig:acs} as an overlay, with the
FUV contours  rendered in  black on the  $V$-band image for  a clearer
sense of the spatial relationship  between the blue star forming knots
(seen in  the FUV) and the  dust complex (seen in  $V$-band).  We have
labeled each blue knot following the scheme in \citet{odea01} to allow
for  easier comparison  of results.   In the  original STIS  NUV image
discussed in  \citet{odea01} (see Fig.~\ref{fig:doubledouble}{\it a}),
knots ``3''  and ``4'' appeared to be  separate, individually resolved
regions  of  emission.  In  our  new  higher  sensitivity and  spatial
resolution  imaging  with ACS  we  detect  these  two regions  as  one
filament of emission that extends $\sim1$\arcsec.  In the interests of
consistency we have nonetheless named  this region with the two labels
``3''  and  ``4''  originally  assigned in  \citet{odea01},  and  will
hereafter refer to  the filament as ``Knot 3+4''  when discussing both
regions as  a whole. As in  \citet{odea01}, we will also  refer to the
``nuclear''  FUV emission,  seen in  Fig.~\ref{fig:hrccontour}  as the
contour  cospatial with the  nucleus in  the underlying  $V$-band.  In
Fig.~\ref{fig:hrccontour}({\it  b}) we overplot  the STIS  S25SRF2 NUV
contours on the same $V$-band image (with colors inverted to highlight
the continuum deficit due to  the outer dust lane).  The blue emission
is morphologically nearly identical in  both the FUV and NUV, although
the higher sensitivity  (and longer exposure time) of  the ACS data is
evident  in the  FUV contours  of  Fig.~\ref{fig:hrccontour}({\it a}),
which  map  lower  surface   brightness  features  than  are  seen  in
Fig.~\ref{fig:hrccontour}({\it b}).   We utilize the  ACS/HRC $U$-band
image in our analysis, though do  not present the image in a figure as
it appears  nearly identical to the  knots in Figs.~\ref{fig:acs}({\it
  b}) and  \ref{fig:hrccontour}.

We present an analysis  of these new
data in the  subsections below, framed in the  context of past results
from \citet{odea01}.


\begin{deluxetable*}{lccccccc}
\tablecaption{Photometry of the Blue Knots in 3C~236}
  \tablewidth{0pc}
  \tablehead{
    \colhead{} &
    \colhead{FUV Flux} & 
    \colhead{$m_{\mathrm{F140LP}}$ (FUV) } &
    \colhead{$m_{\mathrm{FS25SRF2}}$ (NUV) } &
    \colhead{$m_{\mathrm{F330W}}$ ($U$-band) } &
 \colhead{} &
 \colhead{} &
 \colhead{} \\
    \colhead{Source} &
    \colhead{erg s$^{-1}$ cm$^{-2}$} & 
    \colhead{(mag)} &
    \colhead{(mag)} &
    \colhead{(mag)} &
 \colhead{$m_{\mathrm{FUV}} -m_{\mathrm{NUV}}$} &
 \colhead{$m_{\mathrm{FUV}} -m_{U\mathrm{-band}}$} &
 \colhead{$m_{\mathrm{NUV}} - m_{U\mathrm{-band}}$} \\
    \colhead{(1)} & 
\colhead{(2)} & 
\colhead{(3)} & \colhead{(4)} & 
\colhead{(5)} & 
\colhead{(6)} & \colhead{(7)} & \colhead{(8)}}
  \startdata    
\cutinhead{Corrected for Galactic Exctinction}                                                          
Knot 1  &$8.305\times 10^{-18}$ & $22.204\pm0.025$  & $22.165\pm0.023$  & $22.074\pm0.015$   & $0.039\pm0.034$ & $0.130\pm0.029$  & $0.091\pm0.027$ \\
Knot 2  &$1.771\times 10^{-17}$ & $21.382\pm0.017$  & $21.840\pm0.019$  & $21.731\pm0.013$ &  $-0.459\pm0.025$  & $-0.349\pm0.021$   & $0.109\pm0.023$  \\
Knot 3  &$1.134\times 10^{-17}$ & $21.866\pm0.021$  & $21.865\pm0.020$  & $21.498\pm0.012$  & $0.001\pm0.029$  & $0.368\pm0.024$  & $0.367\pm0.023$  \\
Knot 4  &$1.670\times 10^{-17}$ & $21.427\pm0.017$  & $21.788\pm0.019$ & $21.813\pm0.013$  &  $-0.361\pm0.025$  &  $-0.386\pm0.021$  & $-0.025\pm0.023$  \\
Knot 3+4  &$2.836\times 10^{-17}$ & $20.871\pm0.013$  & $20.961\pm0.013$   & $20.835\pm0.009$ & $-0.090\pm0.018$ & $0.036\pm0.0158$  & $0.126\pm0.016$   \\
Nucleus  &$9.207\times 10^{-18}$ & $22.092\pm0.024$  & $21.613\pm0.017$ & $20.359\pm0.007$  &  $0.480\pm0.029$  & $1.733\pm0.025$   & $1.254\pm0.018$  \\
All & $9.037\times 10^{-17}$  & $19.612\pm0.008$   & $19.703\pm0.007$   & $19.125\pm0.004$  &  $-0.091\pm0.010$ & $0.487\pm0.008$  &  $0.579\pm0.008$ \\
\cutinhead{Corrected for Galactic \& Internal Extinction}
Knot 1  &$4.240\times 10^{-16}$ & $17.934\pm0.025$  & $18.141\pm0.023$  & $19.711\pm0.015$   & $-0.207\pm0.034$ & $-1.777\pm0.029$  & $-1.570\pm0.027$ \\
Knot 2  &$9.044\times 10^{-16}$ & $17.112\pm0.017$  & $17.817\pm0.019$  & $19.369\pm0.013$ &  $-0.705\pm0.025$  & $-2.258\pm0.021$   & $-1.552\pm0.023$  \\
Knot 3  &$5.790\times 10^{-16}$ & $17.596\pm0.021$  & $17.841\pm0.020$  & $19.136\pm0.012$  & $-0.246\pm0.029$  & $-1.540\pm0.024$  & $-1.294\pm0.023$  \\
Knot 4  &$8.674\times 10^{-16}$ & $17.157\pm0.017$  & $17.764\pm0.019$ & $19.451\pm0.013$  &  $-0.607\pm0.025$  &  $-2.294\pm0.021$  & $-1.687\pm0.023$  \\
Knot 3+4  &$1.448\times 10^{-16}$ & $16.601\pm0.013$  & $16.937\pm0.013$   & $18.473\pm0.009$ & $-0.336\pm0.018$ & $-1.872\pm0.0158$  & $-1.535\pm0.016$   \\
Nucleus  &$4.701\times 10^{-16}$ & $17.822\pm0.024$  & $17.589\pm0.017$ & $17.997\pm0.007$  &  $0.232\pm0.029$  & $-0.175\pm0.025$   & $-0.407\pm0.018$  \\
All &$4.614\times 10^{-15}$ &  $15.342\pm0.008$   & $15.680\pm0.007$   & $16.762\pm0.004$  &  $-0.337\pm0.010$ & $-1.420\pm0.008$  &  $-1.082\pm0.008$ \\
\tablecomments{Photometry of  the blue star forming regions
in 3C~236. We  have presented our results corrected  only for galactic
extinction,  and  then again  with  galactic  and internal  extinction
corrections applied. The  internal extinction correction was estimated
from  the Balmer decrement  using SDSS  spectroscopy (see  section 3
and Table \ref{tab:extinction}).
1$\sigma$ uncertainties  have been derived from count  rate statistics.  
The table does not include the systematic errors due to the absolute 
photometric calibration of the instruments, which are 
 5$\%$, 2$\%$, and  5$\%$ for the 
ACS SBC (FUV), ACS HRC ($U$-band), and STIS MAMA (NUV), respectively.
(1) Source name;
(2) measured FUV flux of each blue source from the ACS SBC F140LP image;
(3) apparent brightness of the FUV emission in the {\it HST} VEGAMAG system; 
(4) apparent brightness in the NUV; 
(5) apparent brightness in the $U$-band;
(6) FUV-NUV color; 
(7) FUV-$U$-band color;
(8) NUV-$U$-band color.  
Larger (more positive) color values correspond to redder colors.}
\label{tab:phot}
\end{deluxetable*}

\subsection{The outer lane and inner dusty disk}

In  Fig.~\ref{fig:acs}({\it  a})   a  significant  deficit  of  galaxy
continuum  due to a  dust lane  extends $\sim  10$ kpc  and runs  in a
northeast-to-southwest direction  south of  the nucleus at  a position
angle of $\sim50\deg$.  A  steep brightness and color gradient defines
a ridge north of the outer  lane, giving way to an inner circumnuclear
disk of gas and dust whose major axis is oriented $\sim30\deg$ (offset
$\sim20\deg$ from the outer lane). It is not apparent whether the lane
and disk are disjoint  structures or a continuous, warped distribution
of gas  and dust. \citet{dekoff00} estimated  the mass of  dust in the
complex to be $\sim 10^7$ M$_\odot$, corresponding to a gas mass $\sim
10^9$  M$_\odot$  given  the   standard  gas-to-dust  ratio  from  the
literature \citep{sodroski94}.

\subsection{Properties of the star forming knots}

We have  measured the total flux  from each knot of  star formation in
the SBC FUV image (Fig.~\ref{fig:acs}{\it  b}) as well as the STIS NUV
(Fig.~\ref{fig:hrccontour}{\it b})  and HRC $U$-band  images.  We have
also  measured the  HRC $V$-band  flux for  knots 2,  4, and  the blue
emission associated  with the nucleus. We have  not performed $V$-band
photometry for knots 1 and 3  as they are not clearly detected in this
band (see  Fig.~\ref{fig:acs}{\it a}).  Fluxes were  measured from the
drizzled  images  using  the  \texttt{apphot} package  in  IRAF.   The
default drizzled pixel units for  ACS are in electrons s$^{-1}$, which
scales to counts by a factor of  the gain. As this is corrected for in
the pipeline's calibration stage,  ``counts'' and ``electrons'' can be
considered equivalent  for the purposes  of this paper,  regardless of
the  instrument  being  discussed.   Count  rates were  summed  in  an
aperture whose radius was chosen based on the size of the source being
measured. Background  count rates measured through an  aperture of the
same radius were subtracted  from the sum. Photometric conversion into
flux  units  was  applied   by  scaling  the  residual  (source  minus
background)  value  by the  inverse  sensitivity  (the {\sc  photflam}
keyword), converting  the value from electrons s$^{-1}$  to flux units
in   erg  cm$^{-2}$   s$^{-1}$   \AA$^{-1}$.   Statistical   1$\sigma$
uncertaintes  were  calculated from  the  measured  count rates.   The
absolute photometric  calibration of  the ACS SBC,  ACS HRC,  and STIS
MAMA is  5$\%$, 2$\%$, and  5$\%$, respectively\footnote{Maybhate, A.,
  et al.~2010, ``ACS  Instrument Handbook'', Version  9.0 (Baltimore:
  STScI).}.   Our  photometry was  corrected  for galactic  extinction
using the  scaling relation by  \citet{cardelli89} and a  color excess
$E\left(B-V\right)=0.011$,  as listed  in the  NASA/IPAC Extragalactic
Database (NED).  Summed and  galactic extinction corrected counts were
also  converted to  magnitudes  normalized to  the  {\it HST}  VEGAMAG
system, defined  such that the magnitude  of Vega is zero  in all {\it
  HST} bandpasses.

As estimates on the {\it intrinsic}  color of each knot is required in
our analysis,  and as each star  forming knot is likely  embedded in a
great  deal  of  dust,  we  have also  corrected  all  photometry  for
estimated  internal extinction.   We had  originally proposed  for and
received   STIS  low  dispersion   long-slit  spectroscopy   to  allow
measurement of the Balmer  decrement (H$\alpha$ / H$\beta$ flux ratio)
in a  2\arcsec slit aligned with  the knots, however  a pointing error
rendered  the data  unusable. In  lieu of  this, we  use  the measured
Balmer decrement from \citet{buttiglione09}, who presented line fluxes
for   3C~236  using   available  Sloan   Digital  Sky   Survey  (SDSS)
spectroscopy.

\begin{figure*}
\plottwo{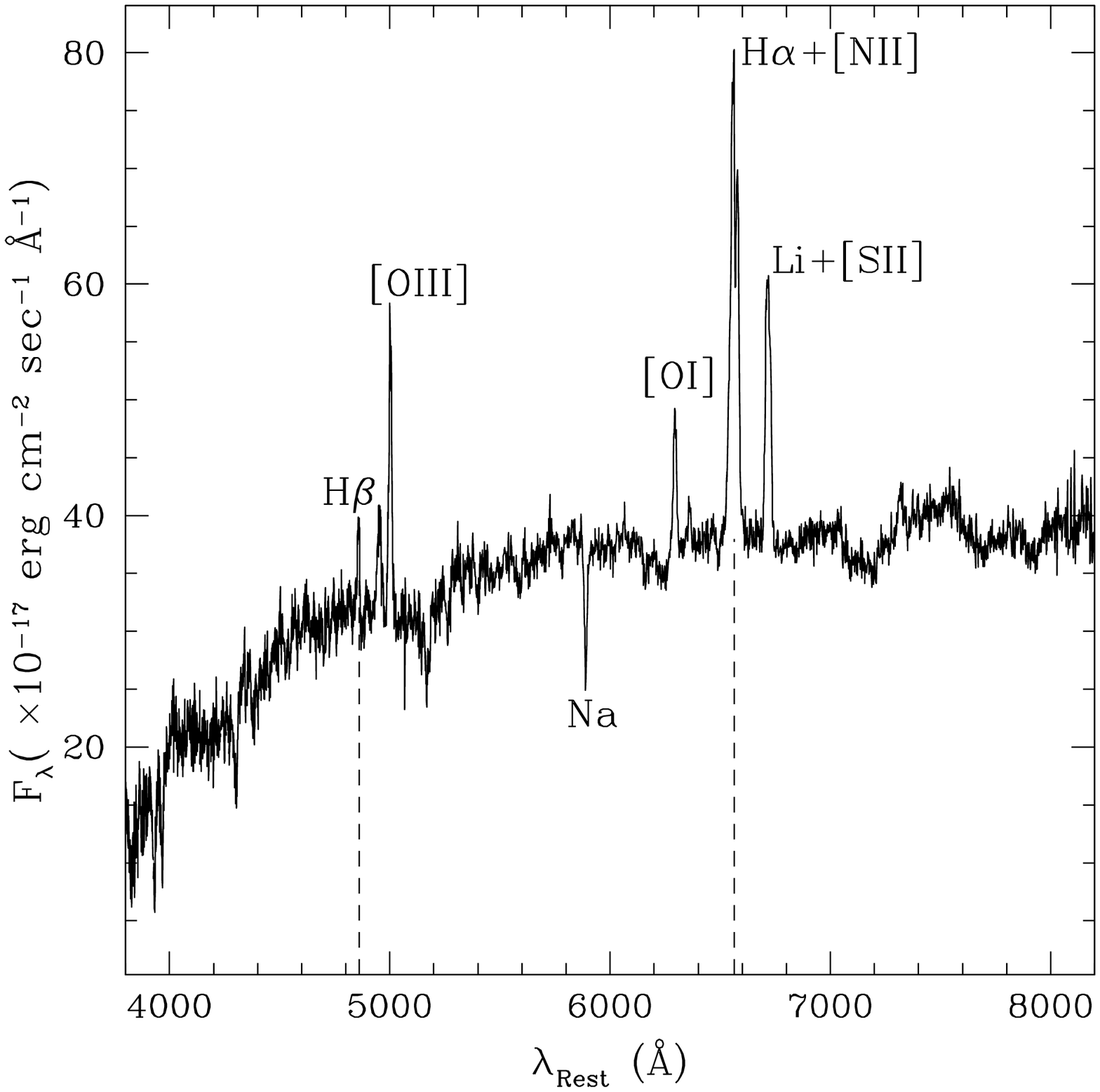}{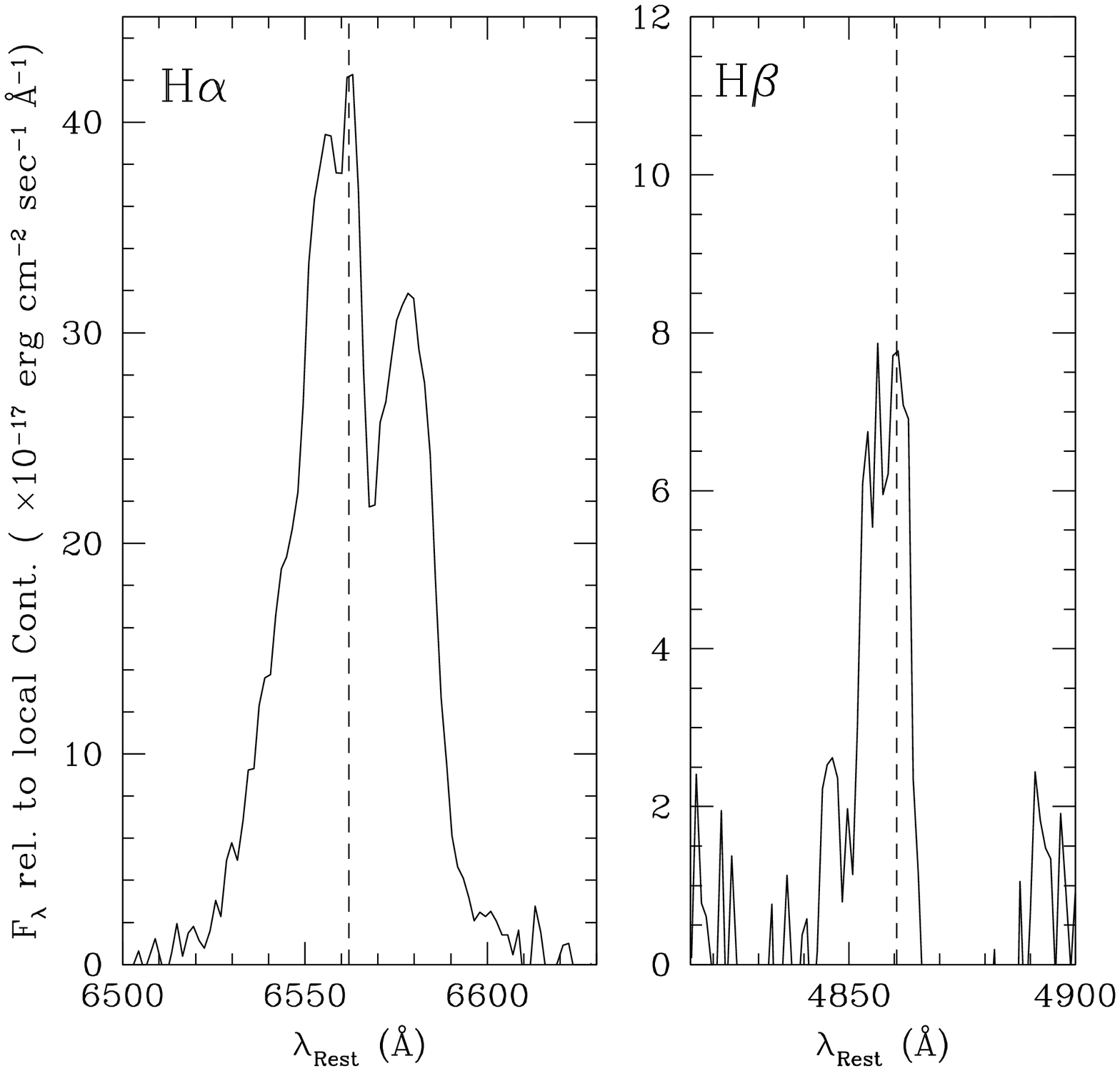}
\caption{({\it left}) Sloan Digital  Sky Survey (SDSS) spectroscopy of
  the  host galaxy  of  3C~236.  Significant  emission and  absorption
  lines  have  been  labeled,  and  the  positions  of  H$\alpha$  and
  H$\beta$, the lines most relevant for our purposes, have been marked
  with vertical  dashed lines.  The  spectrum has been shifted  to the
  rest  frame using  the known  redshift of  3C~236 at  $z=0.1$. ({\it
    right}) ``Zoom in''  of the H$\alpha$ and H$\beta$  lines from the
  spectrum  normalized to  their local  continuum.  SDSS  data  are of
  sufficient spectral resolution to  resolve the two [N{\sc ii}] lines
  at 6548 \AA\ and 6583 \AA\ about H$\alpha$ at 6563 \AA\ (marked with
  a vertical dashed  line).  \citet{buttiglione09} measured the Balmer
  decrement $F_{\mathrm{H}\alpha} / F_{\mathrm{H}\beta} = 4.54$. See Table 4 for other emission line measurements.}
\label{fig:sloan}
\end{figure*}

Assuming  an intrinsic  H$\alpha$/H$\beta$ theoretical  line  ratio of
2.86  (i.e.,   a  ``case   B''  recombination  scenario   is  assumed,
\citealt{osterbrock89}),  the observed Balmer  decrement allows  us to
estimate the  color excess associated with the  internal extinction in
the  source, following the  parameterization of  \citet{cardelli89} as
described in \citet{atek08}:
\begin{equation}
E\left(B-V\right)_{\mathrm{H}\alpha / \mathrm{H}\beta} = \frac{2.5\times \log \left(2.86 / R_{\mathrm{obs}}\right)}{k\left(\lambda_\alpha\right) - k\left(\lambda_\beta\right)}
\end{equation} 
where                        $R_{\mathrm{obs}}                       =
F\left(\mathrm{H}\alpha\right)/F\left(\mathrm{H}\beta\right)$  is  the
observed  flux  ratio, and  the  extinction  curves  at H$\alpha$  and
H$\beta$ wavelengths are $k\left(\lambda_\alpha\right)\approx2.63$ and
$k\left(\lambda_\beta\right)\approx3.71$,  respectively,  as given  by
\citet{cardelli89}.   The  Balmer  decrement  for 3C~236  measured  by
\citet{buttiglione09}   of  $R_{\mathrm{obs}}\approx4.54$   yields  an
$E\left(B-V\right)_{\mathrm{H}\alpha    /   \mathrm{H}\beta}   \approx
0.465$. The  spatial resolution of the SDSS  spectroscopy, while lower
than that of {\it HST}, nonetheless allows for a reasonably believable
estimate   of   the   internal   extinction  in   the   nucleus.    In
Fig.~\ref{fig:sloan}  we present  the  SDSS data  for  3C~236, with  a
``zoom    in''   on    the   H$\alpha$    and   H$\beta$    lines   in
Fig.~\ref{fig:sloan}({\it b}).   As for the purposes of  this paper we
are  primarily interested in  the spectroscopy  solely for  the Balmer
decrement, we  do not fit  any of the  lines and instead use  only the
analysis  from  \citet{buttiglione09}.   We  do  note  that  a  narrow
H$\alpha$  absorption  feature appears  to  be  superimposed over  the
H$\alpha$+[N{\sc ii}] lines, and possibly  offset from the peak in the
emission in velocity space.   H$\alpha$ absorption has been associated
with  hot  stars  \citep{robinson90},  though  we  cannot  confidently
associate the  feature with anything  specific in the nucleus,  as the
spectrum is spatially blended over that size scale.

Moreover,  the lower  spatial  resolution of  SDSS  prohibits us  from
quantifying how patchy the extinction may or may not be on size scales
of  a  few  kpc  (of  order   the  scale  over  which  the  knots  are
distributed).  It  is not unreasonable to  think that one  knot may be
more deeply  embedded in  the dust disk  (and thus  more significantly
reddened) than  another.  Regardless, we expect that  any variation in
reddening between  knots due to  unresolved patchy extinction  will be
reasonably small  when compared to the overall  intrinsic reddening in
the  nucleus. Were  this not  the case,  we would  expect to  see more
severe  color  gradient shifts  in  the  absorption  map presented  in
Fig.~\ref{fig:colormap}.  Instead, the  colors of  the outer  lane and
inner disk appear to be quite uniform over their projected lengths.

In Table  \ref{tab:phot} we present  the results from  our photometric
analysis, including measured FUV flux, magnitudes, and colors for each
knot.   The data  are presented  first with  only  galactic extinction
corrections applied,  and then again  with both galactic  and internal
extinction corrections  applied. These corrections are  given in Table
3.   As the  colors are  always  computed in  magnitudes as  ``shorter
wavelength''$-$``longer  wavelength'', larger  (more  positive) values
correspond to  redder knots.  As  in \citet{odea01}, we find  that the
knots vary relatively significantly in comparing colors that have {\it
  not} been corrected for internal  extinction (columns 6, 7, and 8 in
the  upper  half  of Table  2).   We  note  that  knots  1 and  3  are
significantly redder than 2 and  4. In correcting the colors using our
estimate for internal extinction, however,  we find that all knots are
intrinsically  very blue,  as  expected.  The  sole  exception is  the
UV-bright  emission that  is cospatial  with the  nucleus,  which lies
significantly redward of the ``dust  disk knots'' in color space, even
after the internal extinction  correction has been applied. This could
be due to an older  stellar population or patchy extinction leading to
higher  reddening  along the  line  of  sight  toward the  nuclear  UV
emission (or both).

Again, we  are unable to quantify  whether or not there  may be patchy
extinction on scales finer than  the SDSS spatial resolution. As there
is vastly more dust in the nucleus than in regions of the galaxy a few
kpc outward  from the  dust complex (determined  from our  colormap in
Fig.~\ref{fig:colormap}),   we   find   it   likely   that   we   have
systematically undercorrected for internal extinction. However, a knot
that  is more heavily  extincted than  we have  corrected for  will be
intrinsically bluer and  therefore younger. Uncertainty therefore lies
far more  on the ``young  end'' of age  estimates than it does  on the
old, and we are able to estimate upper limits to the ages of the knots
as done  in \citet{odea01}. We  describe these results in  the section
below.

\begin{deluxetable}{lcc}
\tablecolumns{2}
\tablecaption{Extinction Corrections}
\tablehead{
\colhead{} &
\colhead{Galactic Extinction} &
\colhead{Internal Extinction} \\
\colhead{Band} &
\colhead{$A\left(\lambda\right)_{\mathrm{Gal}}$ (mags)} &
\colhead{$A\left(\lambda\right)_{\mathrm{H}\alpha/\mathrm{H}\beta}$ (mags) } }
\startdata
FUV (F140LP)     & 0.1011 & 4.270 \\
NUV  (F25SRF2)   & 0.0953 & 4.025 \\
$U$-band (F330W) & 0.0559 & 2.362 \\
$V$-band (F555W) & 0.0292 & 1.232 \cr
\enddata
\tablecomments{
Estimated and calculated corrections to photometry 
due to galactic and internal extinction. Galactic extinction 
estimated using $E\left(B-V\right)=0.011$ and the law by \citet{cardelli89}. 
Internal extinction calculated from the Balmer decrement $F\left(\mathrm{H}\alpha\right)/F\left(\mathrm{H}\beta\right)$ as measured by \citet{buttiglione09}
from SDSS spectroscopy, using the method described in \citet{osterbrock89} and in \S3.3.}
\label{tab:extinction}
\end{deluxetable}

\subsection{Comparison of photometry with stellar population models}

We  have compared  our photometry  with evolutionary  models  from the
stellar         population         synthesis         code         {\sc
  starburst99}\footnote{http://www.stsci.edu/science/starburst99/}
\citep{leitherer99,vazquez05}.    The   package  incorporates   Geneva
evolutionary tracks as well  as Padova asymptotic giant branch stellar
models, and is now widely used for its effectiveness in accounting for
all  stellar  phases  contributing  to  the SED  of  a  young  stellar
population from the FUV to the NIR. Several simulations were run, with
parametric variations  in stellar  initial mass function  (IMF), heavy
element  abundance,  mass range,  and  whether  or  not the  starburst
continuously formed stars  at a star formation rate  (SFR), or whether
it  was instantaneous  (wherein the  starburst is  modeled by  a delta
function,  and the  resulting  population is  allowed  to age  through
time).

Simulation results were $K$-corrected  to the redshift of 3C~236 using
its luminosity distance of $\sim 457$ Mpc. The corrected model spectra
were then convolved through the relevant {\it HST} filter transmission
curve     using     the     photometric    synthesis     code     {\sc
  synphot}\footnote{http://www.stsci.edu/hst/HST\_overview/documents/synphot/}
in the STSDAS IRAF package. These corrected and convolved models could
then be directly compared with our extinction-corrected photometry.

An inherent  challenge in comparing our data  to predictive population
synthesis models lies in the fact that we cannot know for certain what
the absolute  intrinsic color  of each star  forming knot is.   As our
internal extinction correction is based solely on the Balmer decrement
as   measured   from   (relatively)   low  spatial   resolution   SDSS
spectroscopy,  we are  able to  at best  make a  rough  correction for
internal reddening.  As such,  running several simulations with slight
variations in model parameters  was largely an exploratory exercise in
which we qualitatively  characterized how much a model  might shift in
color-color space  given slight  changes in, for  example, the  IMF or
heavy element  abundance.  In general, these variations  were found to
be small enough such that  our uncertain photometry would be unable to
discriminate between subtly different  models. 

Ultimately, however, we  are most interested in the  rough ages of the
knots as this study focuses  on a possible coeval relationship between
the blue star forming regions  and the reignition of radio activity in
the   central   engine.    Model   variations  are   such   that   our
order-of-magnitude  age estimates  are largely  independent  of slight
changes  in  model parameters.   Moreover,  slight  variations in  our
estimate for the internal extinction would not affect the shape of the
observed SED so much that we'd associate it with a significantly older
or younger  starburst. As  such, we will  focus our discussion  on the
``best fit''  model we  have found, with  the caveat that  all results
presented in this section are  uncertain and highly dependent upon the
intrinsic SEDs of the knots, which we can only roughly estimate.

It is also  important to note that there is  line contamination in our
broad-band filters that we have not corrected for, and it is necessary
to quantify how  much this might affect our  results, particularly for
the  FUV emission  in the  nucleus.  In  Table \ref{tab:tab4}  we list
intensities of selected bright  emission lines (both directly measured
from  SDSS spectroscopy  as analyzed  by  \citealt{buttiglione09}, and
estimated using  line ratios from \citealt{dopita97},  measured in the
M87 nucleus). In  column (3) of the table we remark  on whether or not
the line falls  within a passband used in  our observations.  Relative
to this  work, there is significant uncertainty  associated with these
line  fluxes given  the  mismatch between  SDSS  resolution and  fiber
placement  and the  blue  star forming  regions  we're interested  in.
Moreover,  there are even  greater uncertainties  in the  estimated UV
line  strengths (as  they are  derived  from ratios  measured for  the
nucleus of  M87, not 3C~236).  We  therefore use these  fluxes only to
roughly  address the  possible  impact of  line  contamination on  our
results.  There are relatively  bright lines contaminating our optical
and UV passbands, namely  [O{\sc iii}]$\lambda5007$ (in ACS HRC F555W)
and  C{\sc iv}$\lambda1549$  (in ACS  SBC F140LP  and  STIS NUV-MAMA).
However,  correcting  for  this  contamination would  not  change  the
overall shape of the SED  relative to the comparison {\sc starburst99}
model  (which itself  includes stellar  and nebular  emission,  and as
mentioned  above,  has  been  redshifted  and  convolved  through  the
appropriate bandpass).  Even correcting the photometry by as much as a
factor of two would not affect the age estimate we'd ultimately end up
making,  both for  the knots  in the  dusty disk  as well  as  for the
nuclear FUV emission.  We therefore  do not correct our photometry for
line  contamination,  given  the  uncertainties mentioned  above,  and
because it ultimately does not affect our results.

In Fig.~\ref{fig:sed}  we present  the main result  of this  paper, in
which we compare  the SEDs of the blue star  forming knots as measured
in our  data with the  synthetic SEDs generated by  {\sc starburst99}.
We plot four ``snapshot'' epochs  of our best-fit model, normalized to
a  continuous  star  formation   rate  of  1  M$_\odot$  yr$^{-1}$,  a
\citet{salpeter55} IMF of slope $\alpha  = 2.35$, upper and lower mass
limits  of 100  M$_\odot$  and 1  M$_\odot$,  respectively, and  solar
abundances  where  $Z=0.020$.  These  ``snapshot''  epochs  are 1  Myr
(lowest line in blue), 10 Myr (second-to-lowest SED in green), 100 Myr
(second highest in  yellow) and 1 Gyr (top SED  in red).  As discussed
previously,  other star  formation  models were  tested,  but are  not
plotted in  Fig.~\ref{fig:sed} as variation  in overall SED  slope and
normalization was minimal.

In black we overplot the measured SEDs of the blue star forming knots,
as well as  the blue emission in the nucleus,  using our photometry as
corrected for galactic and  internal extinction.  Individual knots are
identified  as   per  the  legend   in  the  upper  right   corner  of
Fig.~\ref{fig:sed}, using the naming convention for the knots outlined
in Fig.~\ref{fig:hrccontour}({\it a}).  The  black symbols on each SED
mark the  central wavelengths  of the (from  left to right)  FUV, NUV,
$U$-, and  $V$-band {\it HST} filters, respectively.   As mentioned in
the  previous section,  we  were  only able  to  obtain $V$-band  flux
measurements for knots 2, 4, and the blue emission associated with the
nucleus  (as  they were  the  only  regions  clearly detected  in  the
$V$-band  image).

\subsubsection{The ages of the blue knots cospatial with the dusty disk}

We find the SEDs  of knots 1, 2, 3, and 4 to  best match the $10^7$ yr
model  SED. Still, significant  variation in  color exists  among each
knot, whose  flux in a  particular bandpass can  vary by as much  as a
factor of five with respect  to the others. Moreover, the SEDs compare
somewhat  inconsistently  with  the   model  (green  10  Myr  line  in
Fig.~\ref{fig:sed}), as  there is an  apparent deficit of  emission in
the FUV, an excess in the NUV, and another deficit in the U-band (with
respect to the model).

These variations  are due to  (1) truly intrinsic  differences between
the colors  of the knots,  (2) patchy extinction, and/or  (3) globally
under- or  over-correcting for extinction.  While we  lack the ability
to  quantitatively discriminate  between these  scenarios, we  find it
naturally likely that  a combination of all three  work to produce the
observed  variations.   It is  also  important  to  note that  we  are
studying UV  emission in  a very dusty  environment that  is optically
thick  to UV  photons. The  UV emission  that we  do see  is therefore
likely emitted on the near side of the clouds with respect to our line
of sight,  and we know  little of the  regions of the bursts  that are
more  deeply  embedded  in  the  dusty  disk.   Nevertheless,  we  can
confidently  state  that the  observed  SEDs  of  the knots  are  more
consistent with  a $10^7-10^8$ yr old  starburst than they  are with a
$10^6$ or  $10^9$ yr old  starburst (as modeled by  {\sc starburst99},
anyway).

\subsubsection{The age of the UV emission cospatial with the nucleus}

Along these lines, we find  the nuclear emission to most closely match
the 1 Gyr SED, though there is  a deficit in its FUV flux with respect
to  the  model (top-most  red  1 Gyr  line).   Assuming  that the  FUV
emission  in  the  nucleus is  indeed  associated  with  a 1  Gyr  old
starburst (an assumption  we discuss below), it is  possible (and even
likely)  that  this FUV  deficit  is  due  to an  undercorrection  for
internal  extinction to  the nucleus.   Such an  undercorrection would
affect the  FUV flux  the most, while  the $V$-band flux  would remain
relatively unchanged.   The net  result would be  a steepening  of the
slope on  the blue end,  such that the  SED of the nucleus  might more
closely correspond with  that of the 1 Gyr  model.  Moreover, while we
cannot quantify how patchy the extinction may be over these scales (as
we  have discussed  several  times previously),  it  is reasonable  to
imagine the extinction being greater in  the nucleus than it is on the
edge of  the dusty  disk, given  the inclination of  the disk  is more
edge-on than face-on, such that  the nucleus would be farther down the
line  of   sight  than  the  edge  of   the  disk  \citep{tremblay07}.
Regardless,  we  do not  expect  that  heavier  extinction toward  the
nucleus (in  comparison with  the knots in  the disk) would  alter the
shape of the nuclear SED so  much that it would better correspond with
a different age.

\begin{deluxetable}{lccc}
\tablecolumns{4}
\tablecaption{Emission Line Ratios \& Radio Measurements}
\tablehead{
\colhead{Measurement} &
\colhead{Value} &
\colhead{In Passband?} & 
\colhead{Reference} \\
\colhead{(1)} &
\colhead{(2)} &
\colhead{(3)} & 
\colhead{(4)} }
\startdata
\cutinhead{Measured Intensities}
Log $L$(H$\alpha$)  &  41.13 &  No & 1 \\
H$\beta$ / H$\alpha$  & 0.22(4) & F555W & 1\\
$[$O{\sc iii}$]\lambda$5007 / H$\alpha$ & 0.57(2)& F555W &1\\
$[$O{\sc i}$]\lambda$6364 / H$\alpha$  & 0.30(3) & No &  1\\
$[$N{\sc ii}$]\lambda$6584 / H$\alpha$  & 0.69(1)& No  &1 \\
$[$S{\sc ii}$]\lambda$6716 / H$\alpha$  & 0.49(2)& No  &1 \\
$[$S{\sc ii}$]\lambda$6731 / H$\alpha$  & 0.35(3)& No  &1 \\
Log $L$(178 MHz) & 33.56 & N/A&  2  \\
Log $L$(5 GHz) & 31.62 &  N/A & 3 \\
\cutinhead{Estimated UV Line Intensities}
Ly$\alpha$ / H$\alpha$ & $\sim6.7$ & No \\ 
C{\sc iv} $\lambda1549$ / H$\alpha$ & $\sim1.2$ & F140LP, NUV-MAMA&4\\
He{\sc ii} $\lambda1549$ /  H$\alpha$ & $\sim0.3$ & F140LP, NUV-MAMA & 4 \\
N{\sc iii}$]\lambda1750$ /  H$\alpha$ & $\sim0.1$ & F140LP, NUV-MAMA & 4 \\
C{\sc iii}$]\lambda1909$  / H$\alpha$ & $\sim0.3$ & NUV-MAMA&4\\
Mg{\sc ii} $\lambda 2798$ / H$\alpha$ & $\sim0.1$ & F330W&4 \cr
\enddata
\tablecomments{(1) De-reddened logarithm of luminosity in erg s$^{-1}$ or line ratio (with respect to H$\alpha$); 
(2) measured value, with errors parenthetically presented as percentages;
(3) remark if the line (at the redshift of 3C~236) falls into one of the {\it HST} passbands
used in this study;
(4) reference. UV line intensities were estimated 
using measured ratios from the nucleus of M87. Significant uncertainties
are associated with these estimates. }
\tablerefs{(1) \citet{buttiglione09}; (2) \citet{spinrad85}; (3) 
\citet{buttiglione10}; (4) \citet{dopita97}}
\label{tab:tab4}
\end{deluxetable}

Of course, it is important to estimate how much the AGN may contribute
to the FUV  emission in the nucleus. 3C~236  is a low-excitation radio
galaxy (LEG) that  lacks broad lines \citep{buttiglione09}, suggesting
that the accretion  region is obscured along our  line of sight (i.e.,
\citealt{urry95}).   Contribution  from the  AGN  to  the nuclear  FUV
emission would therefore come in the form of scattered light, which we
do not  expect to  contribute significantly to  the overall  UV excess
(certainly not  to a degree  that would affect  our order-of-magnitude
age  estimates).   \citet{holt07}  similarly  fit  stellar  population
models to their  ground-based SED for 3C~236, finding  that a good fit
required a red  power-law component, suggestive that the  UV excess is
not likely associated with scattered light from the AGN and is instead
a  young, reddened stellar  population. Moreover,  no point  source is
visible  in any {\it  HST} imaging  of the  nucleus, and  {\it Spitzer
  Space Telescope} IR spectroscopy  obtained for 3C~236 ({\it Spitzer}
programs 20719 and  40453 by PI Baum and  collaborators) does not show
evidence for  a hidden quasar continuum  (although formal presentation
of these data is forthcoming  in future papers).  Hence, we expect the
majority  of the  FUV emission  in the  nucleus to  be due  to stellar
continuum from  young stars, whose age  we have estimated  to be $\sim
10^9$ yr old.

\subsubsection{The ages of the knots in context of other works}

\begin{figure*}
\plotone{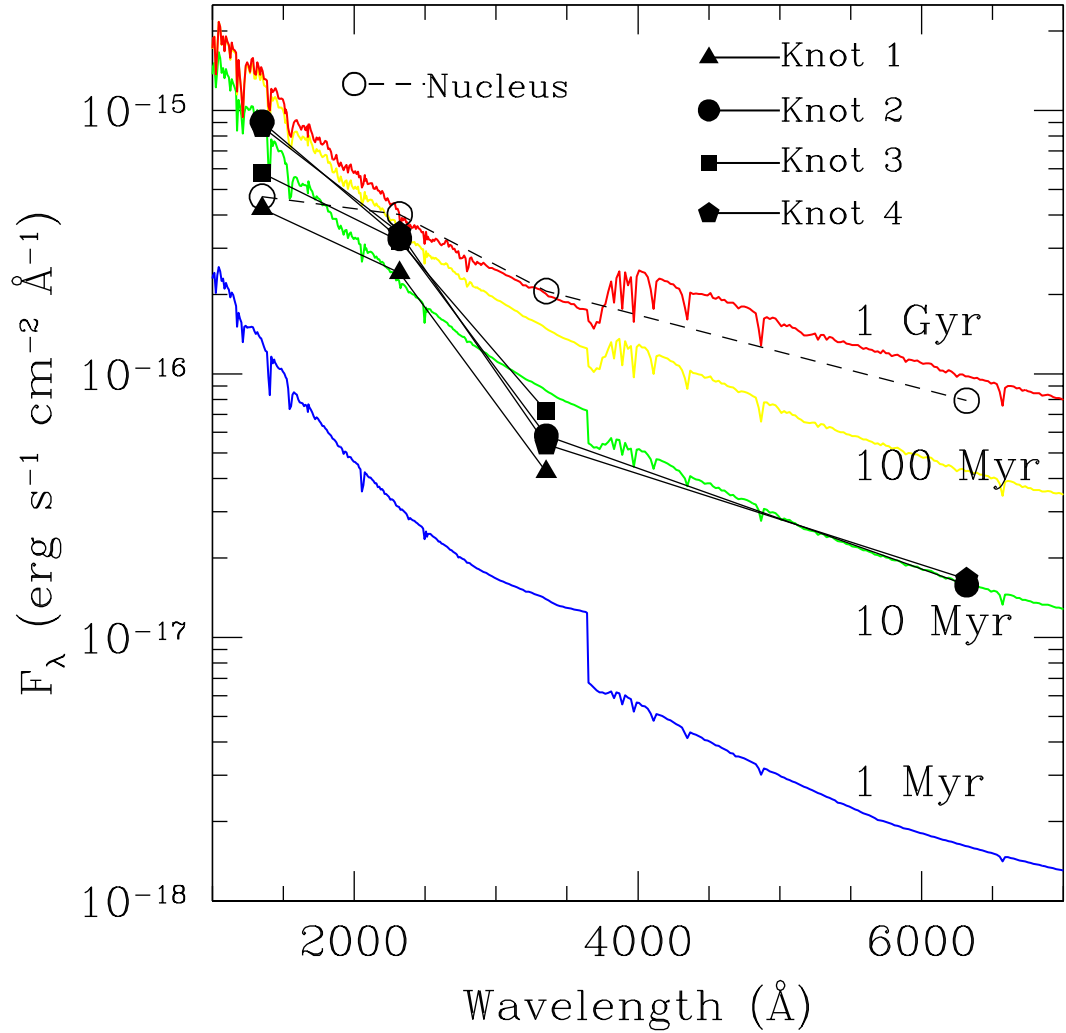}
\caption{SEDs of the blue star  forming knots as measured in our data,
  compared  with  synthetic SEDs  as  predicted  by {\sc  starburst99}
  stellar  population  synthesis models.   We  plot four  ``snapshot''
  epochs  of one  standard {\sc  starburst99} model,  normalized  to a
  continuous star formation rate  of 1 M$_\odot$ yr$^{-1}$, a Salpeter
  IMF of  slope $\alpha =  2.35$, upper and  lower mass limits  of 100
  M$_\odot$ and 1 M$_\odot$,  respectively, and solar abundances where
  $Z=0.020$. Other  continuous star formation models  were tested, but
  are  not  plotted  here  as  variation  in  overall  SED  slope  and
  normalization was  minimal. In black  we overplot the fluxes  of the
  blue  star forming  knots, as  well as  the red  nucleus,  using our
  photometry as corrected for  galactic and internal extinction (using
  the Balmer decrement). Individual knots are identified as per the 
legend in the upper right corner, using the naming convention for the 
knots described in Fig.~\ref{fig:hrccontour}. }
\label{fig:sed}
\end{figure*}

To summarize,  we conclude from Fig.~\ref{fig:sed} that  the four star
forming knots cospatial with the dusty disk are of order $\sim10^7$ yr
old,  while  the stars  in  the  nucleus are  older,  with  an age  of
$\sim10^9$  yr old. In  Table 5  we present  the star  formation rates
associated  with our  ``best fit''  {\sc starburst99}  continuous star
formation  model.   We  also  list  the mass  range  required  for  an
instantaneous  burst model  to  reproduce the  observed fluxes,  while
noting  that   the  continuous  model   is  a  better  match   to  our
data. Nevertheless,  the mass ranges  we estimate are  consistent with
those predicted  by the instantaneous burst  models of \citet{odea01}.
Our estimated  SFRs are  also consistent with  those derived  from the
O'Dea continuous star formation models, and are always on the order of
a few M$_\odot$  yr$^{-1}$.  See Table 3 in  \citet{odea01} to compare
both  their derived  SFRs and  masses with  our results.   

Unlike this work, \citet{odea01}  compared their photometry to Bruzual
\&      Charlot      stellar      population     synthesis      models
\citep{bruzual93,charlot01} absent an estimate for internal extinction
(the  required  data was  not  available  at  the time).   They  found
bimodality  in the  ages of  the  knots, wherein  knots 1  and 3  were
measured to  be young, of  order $\lae10^7$ yr  old, and 2 and  4 were
estimated at  $\lae10^{8}-10^{9}$ yr old.  In contrast,  we estimate a
nearly uniform  age distribution among the  knots in the  dust disk of
$\sim10^7$ yr old, while the FUV  emission in the nucleus is likely to
be $\sim10^{9}$ yr.   The difference in results between  this work and
that of \citet{odea01} likely arises  from (1) different data sets (2)
the lack of internal extinction correction in the O'Dea work (none was
available at  the time),  and (3) comparison  with Bruzual  \& Charlot
models vs.~{\sc starburst99} models  in this work.  Whatever the case,
the age upper limits we estimate for our knots are consistent with the
``young'' knots  1 and 3  from \citet{odea01}. That work  also clearly
emphasizes that the older ages estimated  for knots 2 and 4 are merely
upper limits, and entirely dependent  upon the intrinsic colors of the
knots.

The work by \citet{holt07} also estimated the ages of the star forming
regions  in 3C~236  by fitting  young stellar  population models  to a
ground-based  SED from  the  ESO Very  Large  Telescope (VLT).   Their
results  are  consistent  both  with  \citet{odea01}  and  this  work,
estimating  similar ages  for  the stellar  populations. In  examining
stellar absorption  features, they further  rule out models  with ages
$\gae 1.0$ Gyr,  as those models overpredict the depth  of the Ca {\sc
  ii} feature.

The  work  by  \citet{koekemoer99}  performed  a  similar  age  dating
exercise for  the centrally  dominant radio-loud elliptical  galaxy in
the cooling  flow cluster  Abell 2597. Much  like 3C~236,  that object
possesses a compact radio source, a significant filamentary dust lane,
and a network  of clumpy knots and filaments  of blue continuum, which
\citet{koekemoer99} interpreted to be  sites of recent star formation.
Using single-burst  Bruzual \& Charlot models, that  work derived ages
for the knots  of $\sim 10^7-10^8$ yr, comperable  to the inferred age
of the compact radio source.

\begin{deluxetable}{lcc}
\tablecolumns{3}
\tablecaption{Estimated Star Formation Parameters}
\tablehead{
\colhead{} & 
\colhead{Est. Required SFR} &
\colhead{Est. Required Mass Range} \\
\colhead{Source} &
\colhead{(M$_\odot$ yr$^{-1}$, Cont. Model)} &
\colhead{(Log M$_\odot$, Inst. Model)}\\
\colhead{(1)} &
\colhead{(2)} &
\colhead{(3)}}
\startdata
Knot 1      & 0.69 & $7.3 - 8.7$\\
Knot 2      & 1.47 & $7.7 - 9.0$\\
Knot 3      & 0.94 & $7.5 - 8.8 $\\
Knot 4      & 1.41 & $7.7 - 9.0$\\
Knot 3+4    & 2.36 & $7.9 - 9.2$\\
Nucleus     & 0.47 & $7.4 - 8.7$\\
All         & 7.51 & $8.4 - 9.7$\\
\enddata
\tablecomments{(1) Source identifier; (2) Estimated star formation rates (SFRs) using the ``best fit'' {\sc starburst99} ``snapshot'' epoch as determined based on Fig.~\ref{fig:sed} and as described in \S4.2. We have chosen the 10$^7$ year SED as the ``best fit'' for all of the knots, and the 10$^9$ yr SED for the nucleus. The chosen {\sc  starburst99} model for this estimate is 
normalized  to a continuous star formation rate  of 1 M$_\odot$ yr$^{-1}$, a Salpeter IMF of  slope $\alpha =  2.35$, upper and  lower mass limits  of 100 M$_\odot$ and 1 M$_\odot$,  respectively, and solar abundances where $Z=0.020$. Other models were tested, but those results are not presented here as variation was minimal, and the SFR for each knot is generally of order $\sim1$ M$_{\odot}$ yr$^{-1}$ regardless of the model. (3) Required mass range of an instantaneous 
starburst triggered $10^7 - 10^8$ yr ago. The {\sc  starburst99}  parameters used are the same as in (2), except 
the starburst is modeled as a delta function at time zero, rather than forming continuously with a star formation rate. } 
\label{tab:extinction}
\end{deluxetable}

\begin{figure*}
\plottwo{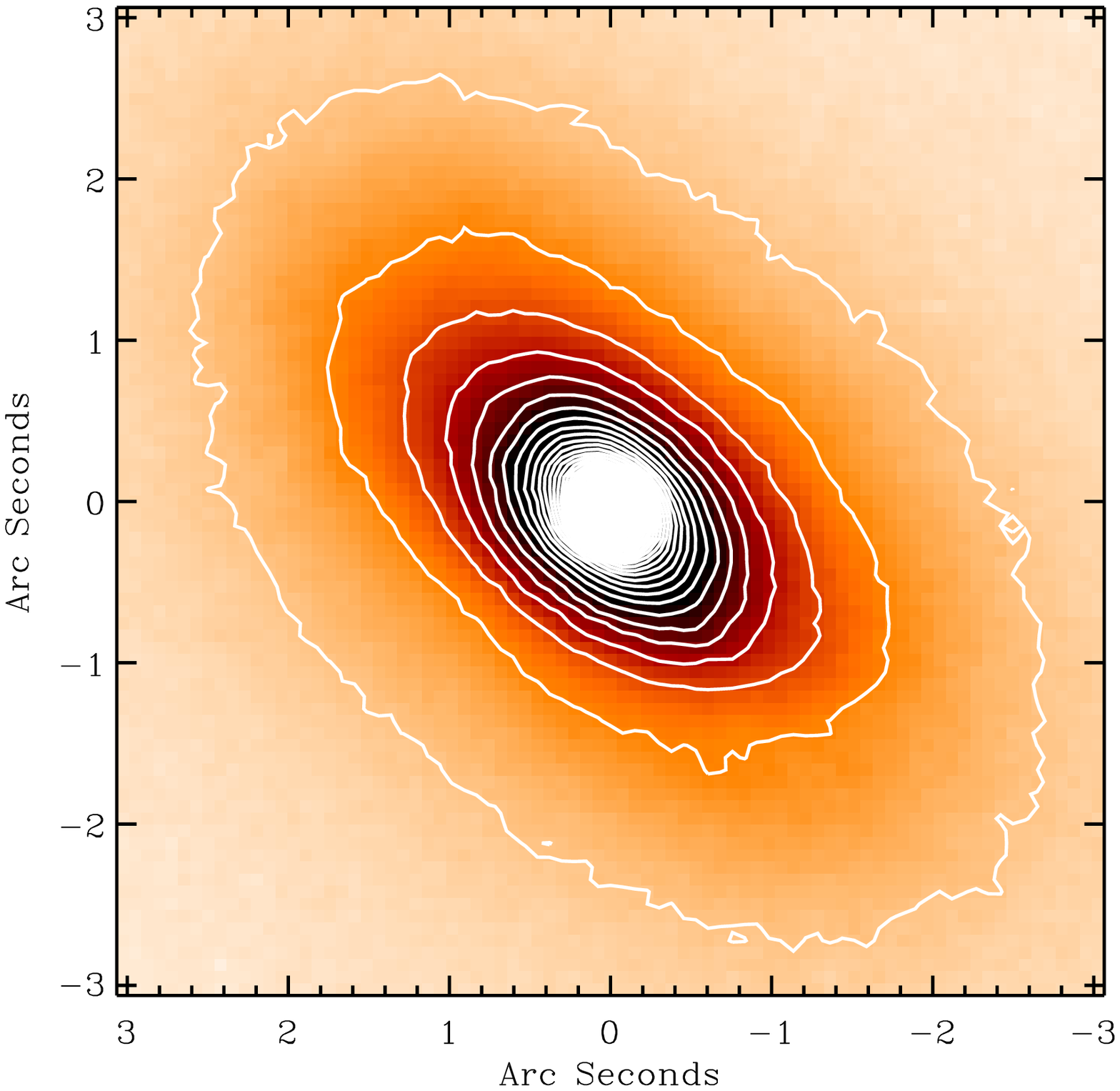}{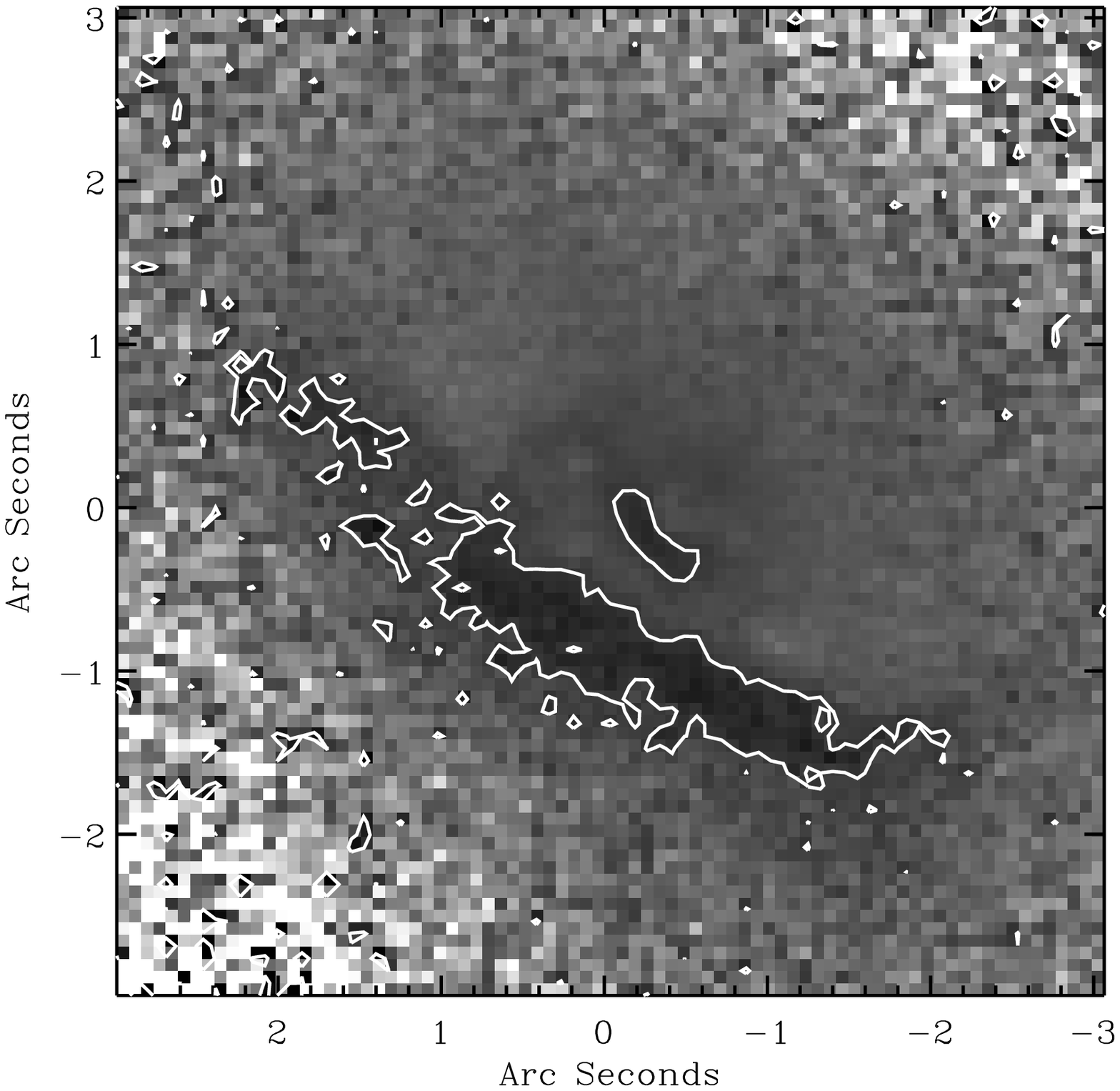}
\plottwo{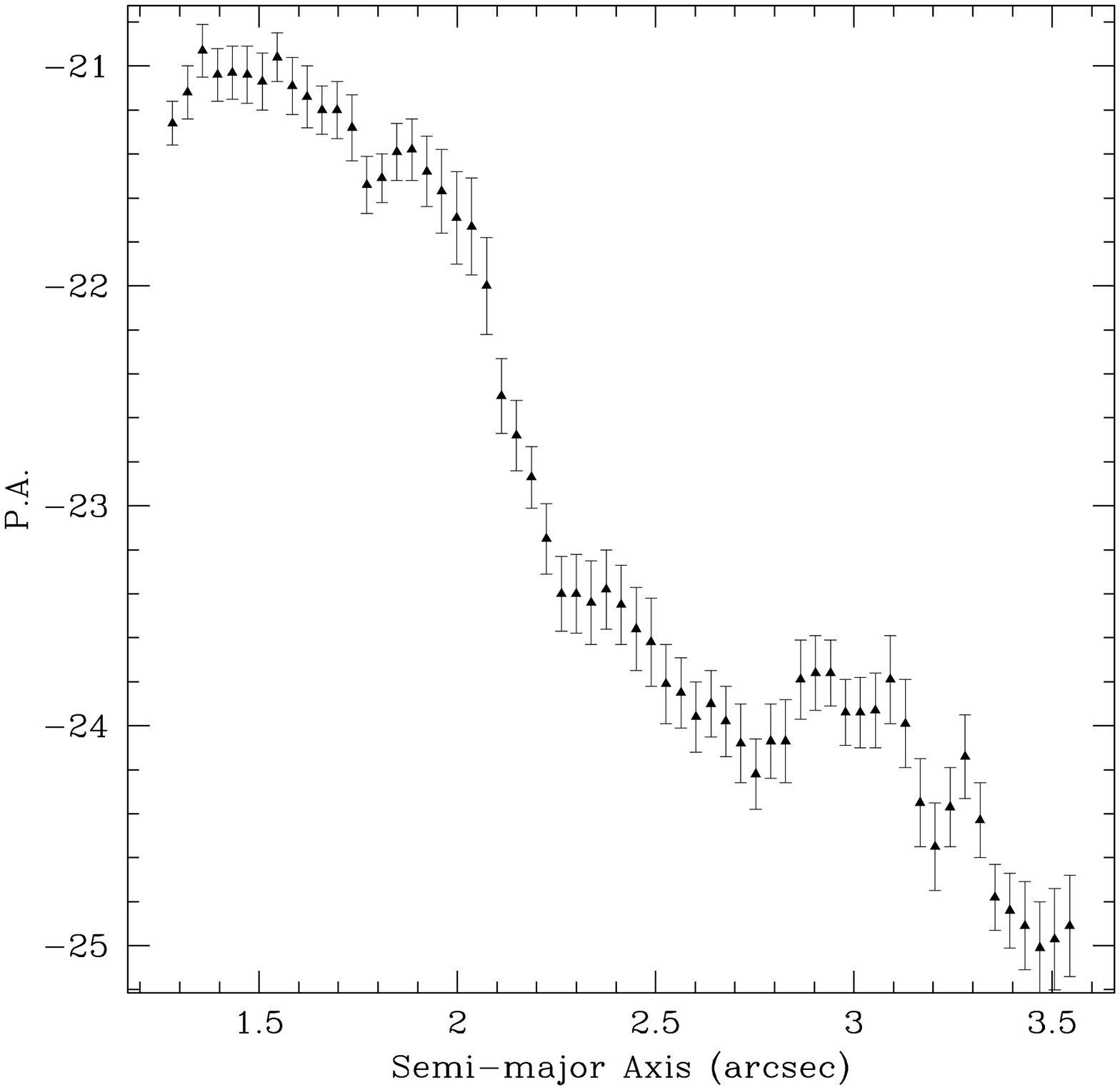}{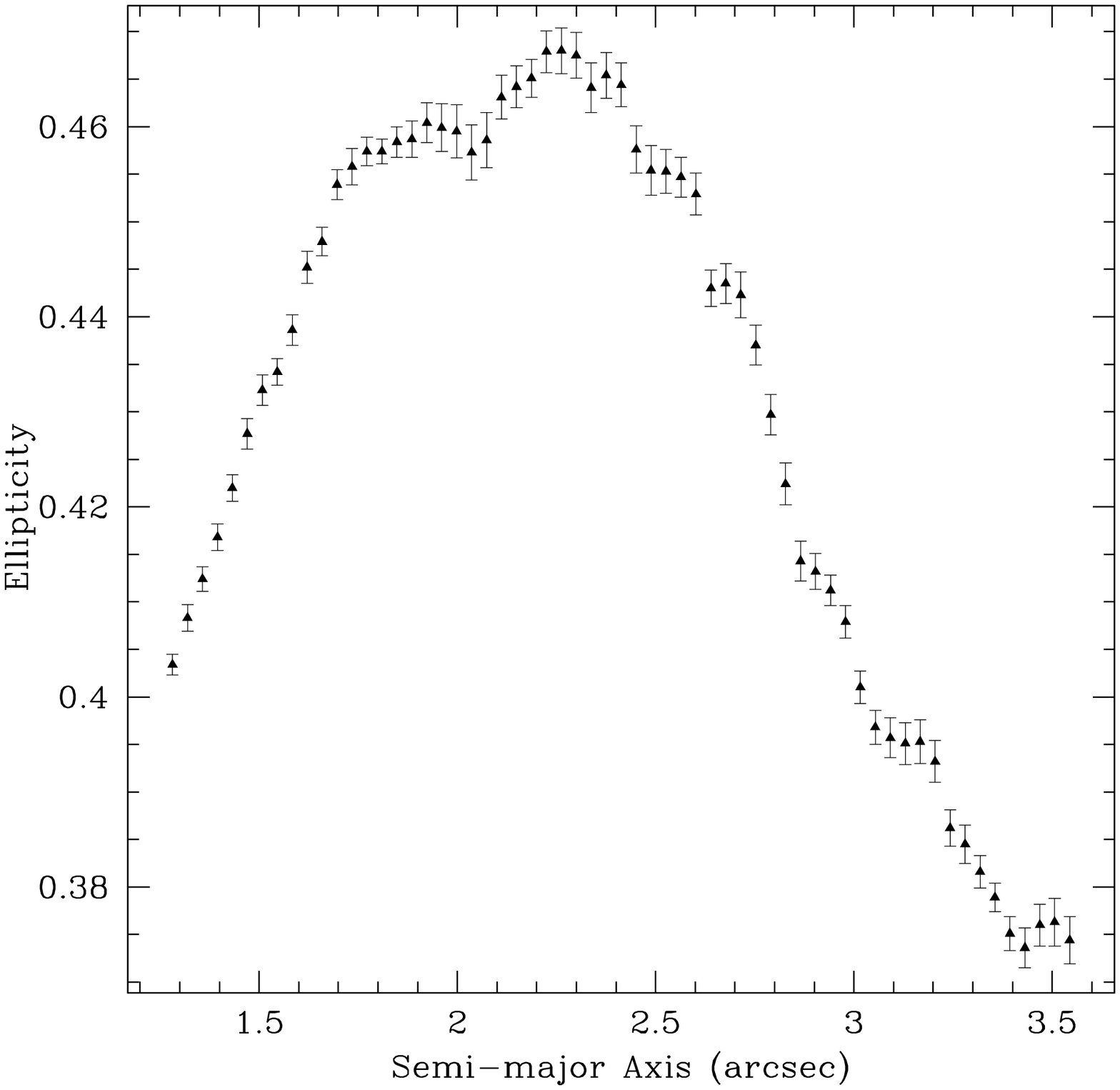}
\caption{({\it top left}) {\it  HST}/NIC2 $1.6$ $\mu$m image of 3C~236
  with highlighted isophotal contours.  ({\it top right}) $1.6$ $\mu$m
  /  0.702 $\mu$m  ($H$/$R$-band) colormap  of the  dusty disk  in the
  nucleus of  3C~236, made  via division of  {\it HST}/NIC2  and WFPC2
  data. Evidence  for two  disparate dust structures  are seen  in the
  colormap. The position  angles of the outer dust  lane and the inner
  disk-like  structure are  offset by  $\sim 10^\circ$.   ({\it bottom
    left}) Semi-major  axis position angles (P.A.) of  ellipses fit to
  the  {\it  HST}/NIC2  $H$-band  isophotes  using  the  IRAF  routine
  \texttt{ellipse}.   The plot  shows  a several  sigma  twist in  the
  isophotes  between  1\arcsec and  3\arcsec,  which  is also  visibly
  evident  in the contours  of the  image at  top left.   ({\it bottom
    right})  Ellipticity of  the fit  isophotes  from \texttt{ellipse}
  over the same scale.}
\label{fig:nic}
\end{figure*}

\section{Discussion}

\subsection{Dynamics of the gas and dust in 3C~236}

As discussed  in section 3.1 and seen  in Fig.~\ref{fig:acs}({\it a}),
3C~236  possesses both  an outer  filamentary dust  lane and  an inner
circumnuclear  disk  whose semi-major  axes  are  misaligned with  one
another  by $\sim$15$^\circ$  \citep{martel99,dekoff00,odea01}.  These
morphologies  are strongly indicative  of a  galaxy that  has recently
acquired  gas from  a companion,  suggesting  that the  outer lane  is
dynamically  young.  Below  we  motivate this  assertion using  simple
dynamical arguments.

Gas and dust  acquired through mergers or tidal  stripping is expected
to   coalesce   on   a    dynamical   timescale   ($\sim   10^8$   yr,
e.g. \citealt{gunn79,tubbs80})  and precess about a  symmetry plane of
the host galaxy, finally settling into it on a precession timescale of
order a Gyr (e.g.,  \citealt{tohline82,habe85}).  During this time the
gas will dissipate angular momentum and fall inward toward the nucleus
at a rate  dependent upon the structure of the  potential well and the
star formation efficiency of the gas \citep{barnes96,bekki97}.  In the
scenario  proposed by  \citet{lauer05},  filamentary distributions  of
dust  that have not  yet reached  the nucleus  would be  classified as
lanes, which  might be thought  of as transient structures  that would
eventually  form  a  nuclear  disk  if given  sufficient  time  (e.g.,
\citealt{tremblay07}, and references therein).

The  recent study by  \citet{tremblay07} lent  evidence in  support of
this scenario.   That work described  a dichotomy between  dusty lanes
and disks  in a  sample of low-redshift  ($z<0.3$) 3CR  radio galaxies
(including   3C~236),   finding   round   nuclear   dusty   disks   to
preferentially  reside in round  to boxy  host galaxies,  depending on
their  inclination with  respect to  the line  of  sight.  Conversely,
filamentary dust lanes which had not yet settled into the nucleus were
found   to   reside   exclusively   in  host   galaxies   with   disky
isophotes. Numerical  simulations of dissipational  mergers have shown
that rotationally  supported, disky systems are typically  the result of
gas-rich  mergers, while boxy  galaxies are  often formed  through dry
(gas-poor)   mergers   (e.g.,  \citealt{barnes96,bekki97,khochfar05}).
Past studies of both radio-loud and radio-quiet ellipticals have shown
that dust lanes are very often misaligned with the major axis of their
host  galaxy isophotes,  while the  opposite is  true for  dusty disks
(e.g.,  \citealt{tran01},  and  references  therein).   These  results
support a scenario in which nuclear dusty disks are native to the host
galaxy pre-merger, while dust lanes are far younger structures, having
recently been externally acquired  through tidal stripping or a merger
\citep{tremblay07}.

3C~236  appears to contain  {\it both}  an outer  dust lane  and inner
dusty disk. Moreover, the outer  dust lane is slightly misalinged with
the  major  axis  of  the  host  galaxy  isophotes  over  a  projected
4\arcsec\  linear extent  centered  about the  nucleus.   We mark  the
isophotal  structure of  the host  in Fig.~\ref{fig:nic}({\it  a}), in
which we have  plotted isocontours on 1.6 $\mu$m  (analog of $H$-band)
{\it     HST}/NICMOS     imaging     from    \citet{madrid06}.      In
Fig.~\ref{fig:nic}({\it  b}) we present  a ``zoomed  in'' view  of the
colormap   originally  presented   in   Fig.~\ref{fig:colormap},  with
contours marking the misalinged outer lane and inner dusty disk.  Note
how the dusty disk appears to be better aligned with the major axis of
the inner host galaxy isophotes than the lane (Figs.~\ref{fig:nic}{\it
  a}    and   ({\it    b})    are   on    the    same   scale).     In
Figs.~\ref{fig:nic}({\it c}) and ({\it d}) we plot isophote major axis
position angle  (P.A.)  and isophotal ellipticity,  respectively, as a
function of semi-major  axis. These data stem from  fits to the NICMOS
$H$-band  isophotes using the  IRAF task  \texttt{ellipse}, originally
performed  by \citet{donzelli07}  and analyzed  by \citet{tremblay07}.
Note from Fig.~\ref{fig:nic}({\it c}) that the isophotes are lopsided,
particularly  to the  southeast between  $\sim 2$\arcsec  and 4\arcsec
from  the  nucleus.  The  asymmetric  isophotes  are  indicative of  a
stellar population that has yet to fully relax dynamically, suggesting
a recent minor merger has taken place.

Framed in  the context of the  results discussed above,  the dual dust
morphologies  in 3C~236  present strong  evidence of  the  host galaxy
having  recently  ($\sim 10^9$  yr)  acquired  gas  from a  companion.
Following  the  scenario discussed  by  \citet{tremblay07}, the  inner
dusty disk  may have been native  to the host galaxy  {\it before} the
recent  gas  acquision  event,  while  the  outer  lane  may  be  been
externally acquired  as a result  of that event,  and is still  in the
process of migration toward the nucleus.

As  dust  traces molecular  gas  accretion  reservoirs  that fuel  AGN
activity, this scenario works naturally with the notion that 3C~236 is
an ``interrupted''  radio source. \citet{dekoff00}  estimated the mass
of  dust  in   the  lane  and  disk  to   be  $\sim  10^7$  M$_\odot$,
corresponding  to  a  gas  mass  of $\sim  10^9$  M$_\odot$.   Such  a
significant gas mass could supply fuel to the AGN for a long period of
time, allowing  the radio  galaxy to  grow to its  very large  size (4
Mpc). The relic radio source also lends an unrelated argument in favor
of a scenario in which a large  fraction of the gas mass was native to
the host galaxy {\it prior} to whatever event cut off or smothered the
AGN  activity  for  a  period  of  time  (ostensibly  the  same  event
responsible  for  depositing new  gas  into  the  system). {\it  Some}
significant fuel  source was clearly  necessary to power such  a large
radio  jet, though it  is not  clear whether  that original  source is
still present and  observed in the form of the  inner disk, or whether
the original source was exhausted and the dust we observe now traces a
newly deposited  accretion reservoir. Imagining for a  moment that the
inner dust disk is old and native to the host ``pre-interruption'', it
is plausible  to imagine the  inner disk being perturbed,  cutting off
the BH  fuel supply, then resettling  back toward the  nucleus after a
short period of time ($\sim 10^7$ yr), reigniting AGN activity. During
this  time the young,  newly acquired  dust from  the ``interruption''
event  would  begin to  settle  into  a  filamentary lane,  and  begin
migration inwards toward the nucleus on a much larger timescale (a few
times $\sim 10^9$ yr).

The  nature of  this ``interruption''  event is  not clear.   The host
galaxy of  3C~236 is  in a  very poor environment  and has  no obvious
group members  within 0.5 Mpc  (based on a statistical  correction for
background  contamination, \citealt{zirbel97}).   This,  combined with
the fact that the isophotes  of 3C~236, while lopsided, are not highly
irregular, suggests that the event was likely a minor merger.  In this
scenario, the  presumably dust-rich and small donor  galaxy would have
been fully consumed by the much larger 3C~236.  Were the event a major
merger, one would expect to see more irregularity in the isophotes, as
the ``dynamically  young'' dust,  the CSS radio  source, and  the fact
that  one  can  still  detect   the  relic  radio  source  even  given
synchrotron  cooling  timescales  ($\sim  10^8-10^9$ yr)  require  the
merger  event  to have  happened  relatively  recently  (no more  than
$10^7-10^8$  yrs ago). A  major merger  of nearly  equal-mass galaxies
would likely result in  structural irregularities that would last much
longer than that timescale.

\subsection{Star formation in the nucleus in the context of the 
relic radio source}

Canonical  estimates  based on  synchrotron  cooling timescales  would
suggest that  the relic $\sim 4$  Mpc FR~II radio source  is not older
than a  few times $10^8$ yr  (e.g., \citealt{parma99}). \citet{odea01}
attempted to  better constrain  the age of  the relic  source, arguing
that  a  lower limit  to  its  age  $t_{\mathrm{min,relic}}$ could  be
estimated using simple dynamical arguments: 
\begin{equation}
t_{\mathrm{min, relic}} \simeq 7.8 \times 10^6 \left({v_{\mathrm{lobe}} \over
  c}\right)^{-1} \ \ {\rm yr}
\end{equation}
where  $v_{lobe}$ is the  lobe propagation  speed.  Using  a canonical
expansion  speed  of  $\sim  0.03  c$  (e.g.,  \citealt{alexander87}),
\citet{odea01} estimated the  dynamical age of the relic  source to be
$2.6\times 10^8$ yr.  This would make for an old radio source, near to
the time when synchrotron  cooling would render the lobes unobservable
at  higher frequencies  as the  electrons age.   Of course,  the relic
source associated with 3C~236 is among the largest in the universe, so
the fact that it's likely old is not surprising.

If the double-double radio source indeed arises from episodic activity
in the AGN, at some stage in the life of 3C~236 its nucleus would have
entered a dormant phase and  halted the collimation of its jets.  This
would  have  deprived  the  hot-spots  of their  energy  supply  on  a
timescale of  order the nucleus-to-hotspot  traversal time of  the jet
material  ejected prior  to  the shut  down  of the  AGN, assuming  it
advanced                    relativistically                    (e.g.,
\citealt{baum90,kaiser00,schoenmakers00a,schoenmakers00b}).         The
production    of   ``young    electrons''   in    the    hotspots   by
magnetohydrodynamical  (MHD) turbulence or  Fermi acceleration  in the
Mach disk is thought  to cease when the hotspot is no  longer fed by a
jet, meaning the  electron population will begin to  age once the last
of the  remaining jet  material has arrived  (e.g., \citealt{jones99},
and  references  therein).  If  one  believes  radio  spectral  ageing
techniques, then the ages of  the youngest electrons in the lobe added
to the jet traversal time from the nucleus to the hotspot should be of
order   the    timescale   over    which   the   nucleus    has   been
dormant. \citet{odea01} estimated this timescale to be $\sim 10^7$ yr,
an order  of magnitude younger  than the $2.6\times10^8$  yr dynamical
age of  the relic source.

Note that the  ages estimated by these two  techniques need not agree,
as the ages  of the youngest electrons in the  lobe will correspond to
the time  when jet propagation ceased,  but tell us  nothing about how
long the  nucleus may  have been active.   Moreover, all of  these age
estimates in the radio are  heavily dependent upon assumptions such as
the propagation speed and the true source of young electrons.  Indeed,
the  active phase  corresponding to  the creation  of the  relic radio
source may  have been far longer  lived than the dynamical  age of the
relic source itself.  We cannot know for certain how much of the lobes
may have already cooled  past the point of observability. Observations
at lower frequencies may be enlightening in this regard.

As  discussed in  section  3.3.2, we  have  estimated the  age of  the
nuclear  starburst to  be $\sim  10^9$  yr old,  given the  previously
discussed  caveats  and noting  that  it may  be  younger  if we  have
undercorrected for internal  extinction to the nucleus.  Nevertheless,
the approximate  ages of  the nuclear starburst  and the $\sim  4$ Mpc
relic source are  just close enough to warrant  noting that they might
{\it possibly}  be related to a  common gas infall  event. Having said
that, we  again stress  that this  is only a  possibility, and  we are
unable to draw  conclusions relating to such a  connection in light of
the significant uncertainties involved.

\subsection{The star forming knots in the context of the CSS radio source}

\citet{odea01} used arguments similar  to those discussed in the above
section to estimate the age of the central CSS radio source, operating
under the assumption that it  was young and not ``old and frustrated''
(a reasonable assumption, especially in light of recent results on CSS
sources,  e.g.    \citealt{holt09}).   Their  estimate   for  its  age
$t_{\mathrm{min,CSS}}$ was given by
\begin{equation}
t_{\mathrm{min,CSS}} \simeq 3.2 \times 10^3 \left({v_{\mathrm{lobe}} \over c}\right)^{-1} \ \ 
{\rm yr},
\end{equation}
where using the same lobe advance  speed they found the age of the CSS
source to be  very young indeed, of order  $\sim1.0\times10^5$ yr.  As
discussed  in section  3.3, we  have estimated  that the  star forming
knots cospatial with the dusty disk  are $\sim 10^7$ yr old.

\subsection{Is the recent episode of star formation coupled to the 
rebirth of radio activity in 3C~236?}

If  the recent episode  of star  formation in  the disk  were directly
related to the event resulting  in reignition of radio activity giving
birth to the CSS source,  we would not necessarily expect a one-to-one
correspondence among their ages. The dynamical time on scales of a few
kpc, where the star forming knots  are located, is far longer than the
dynamical time  on sub-pc  scales where the  AGN is fuelled.   In this
context, the  difference in estimated  ages of the star  forming knots
and the CSS source would constrain the timescale over which the gas is
transported from kpc to sub-kpc  scales. This can't be any longer than
the age  of the  young stars ($\sim  10^7$ yr), which  corresponds not
only  with the dynamical  time on  kpc scales  but also  the estimated
dormancy period of the nucleus.

The recent work by  \citet{wuyts09} discusses the merger-driven models
of  \citet{hopkins06} in the  context of  internal color  structure in
massive star forming galaxies.   They find internal color gradients to
be  strongest  during  the  merger  phase, with  blue  star  formation
expected on scales  outside a few kpc (approximately  where we observe
the star  formation in 3C~236).   In their previously  mentioned work,
\citet{koekemoer99} derived  ages to star forming knots  in Abell 2597
that  were closely  related to  the inferred  merger and  AGN fuelling
timescales giving rise to the compact radio source in that galaxy.

Our observations have yielded similar results, and seem to suggest one
possible  ``story'' for  the history  of 3C~236.   In  considering its
double-double radio morphology in the context of its dynamically young
dust  complex and  recently triggered  compact starbursts,  we suggest
that 3C~236  has undergone  multiple epochs of  AGN activity due  to a
non-steady supply of fuel to  the central engine.  We suggest that the
period of  activity related to the  4 Mpc relic source  was ended when
the  fuel  supply  to the  central  engine  was  cut off,  whether  by
exhaustion,  strangulation,   or  disturbance.   After   a  subsequent
$\sim10^7$ yr dormant  phase, infalling gas from a  minor merger event
reached kpc  scales, where a  starburst was triggered  via cloud-cloud
collisions  amid collapse.   The gas  not involved  in  star formation
reached  the nucleus  after  a subsequent  dynamical time,  triggering
reignition of the AGN and giving birth to the CSS radio source.

We hesitate to make further suggestions relating to the star formation
in the nucleus and its  possible connection with the relic source, nor
will we  suggest whether  or not the  outer filamentary dust  lane and
inner disk are two distinct structures with different histories (i.e.,
recently acquired vs.  native to the host pre-merger).  We have argued
in section 4.1 that discriminating  between the two scenarios may shed
light on whether the AGN ``interruption'' event was due to actual {\it
  exhaustion} of its fuel  supply, or only dynamical {\it disturbance}
of its  fuel supply. Whatever the  case, the results of  this work and
those of \citet{odea01} strongly suggest  that the transport of gas to
the nucleus of  3C~236 has been significantly nonsteady  over the past
Gyr, giving  rise to a  unique galaxy that  acts as an  important test
case in studies of the AGN/starburst connection.

\section{Summary and Concluding Remarks}

We have presented follow-up {\it HST} ACS and STIS observations of the
radio galaxy 3C~236, described by \citet{odea01} as an ``interrupted''
radio source.  The galaxy is  associated with a massive relic $\sim 4$
Mpc FR~II  radio source (making it  one of the largest  objects in the
universe), as well as an inner 2 kpc CSS ``young'' radio source.  This
``double-double'' radio morphology is  evidence for multiple epochs of
AGN  activity, wherein  the BH  fuel supply  is thought  to  have been
exhausted or cut  off at some time in the past,  and has only recently
been reignited.

We present {\it HST} FUV, NUV, $U$-, and $V$-band imaging of four star
forming  knots,  previously  described  by  \citet{odea01},  that  are
arranged in an arc along  the outer edge of the galaxy's circumnuclear
dust  disk   (which  itself  is  surrounded  by   a  misaligned  outer
filamentary dust lane). We  have also detected blue emission cospatial
with the nucleus itself. We  describe these observations in detail, as
well as the steps taken to  reduce the data.  We present photometry of
the  blue knots,  and  discuss our  efforts  to correct  the data  for
internal extinction to the source using the Balmer decrement available
from archival SDSS spectroscopy.

We compare the  measured four-color SEDs of the  star forming knots to
synthetic  SEDs from  {\sc starburst99}  stellar  population synthesis
models, with the  ultimate goal of roughly estimating  the ages of the
knots. We  find that the four  knots cospatial with the  outer edge of
the dusty  disk are likely $\sim10^7$  yr old, while  the FUV emission
cospatial with the nucleus is likely older, at $\sim10^9$ yr old (with
the caveat that undercorrection for internal extinction in the nucleus
would lower this limit). We argue that the ages of the young knots are
suggestive of a causal connection with the young central radio source.

We  frame these  results in  the context  of 3C~236  as  an apparently
``interrupted''  radio galaxy.  Our  results are  generally consistent
with those  of \citet{odea01}, and  we argue along similar  lines that
the transport  of gas in the  nucleus of 3C~236  is nonsteady, wherein
the active phase giving rise to the  4 Mpc relic source was cut off by
exhaustion or disturbance of the AGN fuel supply. We suggest that this
lead to a  dormant period punctuated by a minor  merger event, and the
subsequent  infalling gas  triggered not  only a  new episode  of star
formation, but also ushered the  galaxy into a new active phase giving
rise to the young CSS radio source.

Results such as these support  the natural argument that infalling gas
largely fosters  the relationship  between the growth  of the  AGN and
that of its host galaxy  stellar component. In time, that relationship
may evolve from  mutual growth to the regulation of  the latter by the
former through  quenching from  AGN feedback.  As  infalling gas  is a
critical   component  in   merger-driven   hierarchical  models,   the
triggering of AGN  may be fundamental to galaxy  evolution itself.  In
this  way, the  relevance of  3C~236  might extend  beyond studies  of
episodic activity in  radio galaxies to studies of  AGN in the context
of galaxy evolution as a whole.

\acknowledgments   We  thank   Rafaella  Morganti,   Clive  Tadhunter,
Alessandro  Capetti, and  Andy Robinson  for helpful  discussions.  We
also thank Ari  Laor and the anonymous referee  whose helpful comments
lead to the improvement of this work.  G.~R.~T.~acknowledges B.~F.~and
support from the National  Aeronautics and Space Administration (NASA)
grant  HST-GO-9897, as  well as  the NASA/NY  Space  Grant Consortium.
This work is based on  observations made with the NASA/ESA {\it Hubble
  Space Telescope}, obtained at the Space Telescope Science Institute,
which is operated  by the Association of Universities  for Research in
Astronomy, Inc., under NASA  contract 5-26555.  This research has made
use of Sloan Digital Sky Survey (SDSS) data products.  Funding for the
SDSS and SDSS-II was provided  by the Alfred P.  Sloan Foundation, the
Participating Institutions, the  National Science Foundation, the U.S.
Department   of   Energy,   the   National   Aeronautics   and   Space
Administration, the  Japanese Monbukagakusho, the  Max Planck Society,
and the  Higher Education  Funding Council for  England. We  have also
made extensive use of  the NASA Astrophysics Data System bibliographic
services and the NASA/IPAC Extragalactic Database, operated by the Jet
Propulsion  Laboratory,  California  Institute  of  Technology,  under
contract with NASA.

\bibliography{refs}

\end{document}